\documentclass[12pt]{article}%
\usepackage{amsfonts}
\usepackage{amssymb}
\usepackage{amsmath}
\usepackage{enumitem}
\usepackage{geometry}
\usepackage{graphicx}
\usepackage{setspace}
\usepackage{booktabs}
\providecommand{\U}[1]{\protect\rule{.1in}{.1in}}
\newtheorem{assumption}{Assumption}

\usepackage{multirow}
\usepackage{natbib}
\usepackage{xcolor}
\usepackage{hyperref}
\usepackage{bbm}
\newcommand{\cond}{\, \big| \,}
\newcommand{\con}{;}
\newcommand{\potY}[2]{Y_{#2}^{a=#1}}
\newcommand{\obsY}[1]{Y_{#1}}
\newcommand{\true}{\star}
\newcommand{\ind}{\mathbbm{1}}
\newcommand{\AVER}{\mathbbm{P}}

\newcommand{\OR}{{\text{OR}}}
\newcommand{\PS}{{\text{PS}}}

\newcommand{\T}{^\intercal}
\newcommand{\EXP}{E}
\newcommand{\indep}{\rotatebox[origin=c]{90}{$\models$}\,}

\hypersetup{
colorlinks=true,
linkcolor=blue,
filecolor=blue,
urlcolor=blue,
citecolor=blue
}
\usepackage{tikz}
\usetikzlibrary{arrows, arrows.meta, chains, positioning, quotes, shapes.geometric}
\usepackage{tikzit}
\tikzstyle{new style 0}=[fill=white, draw=black, shape=circle, minimum width=35pt]
\tikzstyle{new style 1}=[fill=white, draw=black, shape=circle, dashed, minimum width=35pt]
\tikzstyle{new style 3}=[fill=none, draw=none, shape=circle]

\tikzstyle{new edge style 0}=[-{Latex[length=3mm]}]

\begin{document}

\title{ }

\begin{center}
\textit{Original Article}

{\LARGE Universal Difference-in-Differences\\
for Causal Inference in Epidemiology }

\vspace*{0.1cm}

{\Large Eric J. Tchetgen Tchetgen}$^{\ast}${\Large , Chan Park}$^{\ast}%
${\Large , David B. Richardson}$^{\dagger}$

\vspace*{0.1cm}

* {\small Department of Statistics and Data Science, University of Pennsylvania}

$\dagger$ {\small Department of Environmental \& Occupational Health, University of California, Irvine }

\vspace{0.3cm}

\textbf{Abstract}
\end{center}

\noindent Difference-in-differences is undoubtedly one of the most widely used
methods for evaluating the causal effect of an intervention in observational (i.e., nonrandomized)  settings. The approach is typically used when pre- and post-exposure outcome measurements are available, and one can reasonably assume that the association of the unobserved confounder with the outcome has the same absolute magnitude in the two exposure arms, and is constant over time; a so-called \textit{parallel trends assumption}. The parallel trends assumption may not be credible in many practical settings, including if the outcome is binary, a count, or polytomous, as well as when an uncontrolled confounder exhibits non-additive effects on the distribution of the outcome, even if such effects are constant over time. We introduce an alternative approach that replaces the parallel trends assumption with an \textit{odds ratio equi-confounding assumption} under which an association between treatment and the potential outcome under no-treatment is identified with a well-specified generalized linear model relating the pre-exposure outcome and the exposure. Because the proposed method identifies any causal effect that is conceivably identified in the absence of confounding bias, including nonlinear effects such as quantile treatment effects, the approach is aptly called \textit{Universal Difference-in-differences (UDiD).} Both fully parametric and more robust semiparametric UDiD estimators are described and illustrated in a real-world application concerning the causal effects of a Zika virus outbreak on birth rate in Brazil.

KEY WORDS: Difference-in-differences, equi-confounding, odds ratios, selection
bias, generalized linear models, extended propensity score.

\bigskip\pagebreak

\section{Introduction}

Difference-in-differences (DiD) is a popular approach to account for unmeasured confounding in observational data. DiD is typically used when (i) pre- and post-exposure outcome measurements are available and (ii) the parallel trends (PT) assumption is reasonable, i.e., the absolute magnitude of the association between the unobserved confounder and the outcome is equal across treated and control groups and constant over time. Under PT, an estimate of the causal effect of the treatment can be obtained by taking a difference between treated and control groups of the average change in outcome over time; see \citet{Caniglia2020} for an introduction to DiD and Section \ref{sec:notation} for a brief review. Despite its popularity, PT may not be credible for a variety of reasons; see Section \ref{sec:supp:Limitation PT} of the Appendix. 
Therefore, the development of alternative identifying conditions for DiD settings continues to be an active area of research.

In this paper, we introduce an alternative identification strategy for DiD settings. Specifically, as described in Section \ref{sec:UDiD}, our approach is based on the assumption that confounding bias for the causal effect of interest, defined as an association between exposure and the treatment-free potential outcome, can be identified under a generalized linear model (GLM) relating the pre-exposure outcome and the exposure. The proposed approach allows for investigators to proceed with a new approach in DID settings under a slightly different key identifying assumption (i.e., different from PT) that, if it holds, does not require one to assume equal and additive effects of an uncontrolled confounder under models for binary/polytomous outcomes. An appeal of the framework is that it permits both familiar parametric models, as well as more robust semiparametric estimation approaches, namely (i) a GLM approach, followed by (ii) an extended propensity score approach, and finally, (iii) a doubly robust approach, which remain valid if either (i) or (ii) provides valid inferences without a priori knowing which might be misspecified. Importantly, the proposed methods can be used to identify and estimate the causal effect of a hypothetical intervention in the presence of unmeasured confounding, on virtually any scale of potential interest, including nonlinear scales such as quantile causal effects. For this reason, the approach is aptly called \textit{Universal Difference-in-differences (UDiD).} In Sections \ref{sec:Data} and \ref{sec:Conclusion}, we apply our methods to evaluate the causal link between a Zika virus outbreak and birth rates in Brazil and discuss a sensitivity analysis approach to assess the impact of a violation of key assumptions.


The proposed methods are a special case of a more general approach proposed by \citet{DiD_OREC2022}. This paper serves as an introductory resource on UDiD methods for an epidemiology audience, with a significant emphasis on utilizing parametric models, interpreting identifying assumptions, and offering practical guidelines for implementation and evaluation in practical settings. Specifically, this paper focuses on applying well-established GLMs and propensity score weighting methods to the UDiD framework, thereby easing their adoption in epidemiological studies. Readers interested in more technical details can consult \citet{DiD_OREC2022}.

\section{Notation and A Brief Review of DiD} \label{sec:notation}

Consider a study in which pre- and post-exposure outcomes $\obsY{0}$ and $\obsY{1}$ are observed; let $Y_t^a$  denote the potential outcome at time $t \in \{0,1\}$ under a hypothetical intervention that sets a binary exposure/treatment $A$ to a value $a\in\left\{  0,1\right\}$. 
Following a standard DiD model \citep{CiC2006}, suppose that the treatment-free potential outcome is generated from the following model for $t=0,1$:
\begin{align}
&
\potY{0}{t} = h(U_t,t)
\ , \ 
\quad \quad
h(u,t) = u + \beta_T t
\ , \
\quad \quad
U_t = \beta_0 + \beta_A A + \epsilon_t 
\ , \
\label{eq-standard DiD}
\tag{DiD Model}
\\
\nonumber
&
\epsilon_t \text{ satisfies either }
\bigg\{
\begin{array}{ll}
\text{time independence: } 
& \epsilon_1 | A \stackrel{D}{=} \epsilon_0 | A
\quad \text{ or }
\\[-0.25cm]
\text{treatment independence: } 
&
\epsilon_t | (A=0) \stackrel{D}{=} \epsilon_t | (A=1) 
\end{array}
\ .
\end{align}
Here, $\epsilon_t$ is an unobserved error at time $t$ that is independent of time or treatment, and $U_t$ is therefore also unobserved. Note that allowing $U_t$ to depend on $t$ accommodates the factors predicting $Y_0$ to be distinct from those predicting $Y_1$. In \eqref{eq-standard DiD}, $\potY{0}{t}$ is a deterministic function of $U_t$ (in fact a linear function of the latter), but the exposure mechanism of $A$ given $U_t$ is unrestricted. In terms of the distribution of $U_t$, \eqref{eq-standard DiD} assumes that the conditional distribution of $\epsilon_t$ given $A$ is either stable over time given $A$ or independent of $A$ at each time; DiD strictly only requires that $\epsilon_t$ does not depend on $A$ and $t$ in a manner that they interact on the additive scale. Finally, \eqref{eq-standard DiD} implies rank preservation which rules out any additive interaction between $A$ and $U_1$ in causing $Y_1$. 
In Section \ref{sec:UDiD overview}, we present an alternative structural model compatible with \eqref{eq-model OREC}, thus allowing for heterogeneity of the causal effects of $A$ with respect to $U_1$, i.e., $h$ is allowed to be unrestricted.

\eqref{eq-standard DiD} implies that $E ( \potY{0}{t} \cond A=a ) = \beta_0 + \beta_A a + \beta_T t + E(\epsilon_t \cond A=a)$, which further implies the so-called parallel trends (PT) assumption:
\begin{equation}
E\left(  \potY{0}{1}  - \potY{0}{0}  |A=1\right)
=E\left(  \potY{0}{1}  - \potY{0}{0}  |A=0\right)
.
\label{Parallel Trends}
\end{equation}
Expression \eqref{Parallel Trends} states that, on average, the trajectory of the potential outcomes under an intervention that sets the exposure to its control value, is equal between exposed and unexposed groups. Hence, under no unmeasured confounding of the average additive effect of $A$ on $\potY{0}{0}$ and
$\potY{0}{1}$ respectively, both left- and right-handsides of the display above would
be zero. This gives an alternative interpretation
of PT as an assumption of additive equi-confounding
bias, such that the confounding bias for the effect of $A$ on $\potY{0}{1}$ though
not null,  is equal to the confounding bias for the causal effect of $A$ on
$\potY{0}{0}$ on the additive scale \citep{Sofer2016}. Note that the latter is empirically identified under consistency and no causal anticipation assumptions, which we now state:
\renewcommand{\theassumption}{1a}
\begin{assumption} \label{assumption:consistency}
\textit{Consistency}: $Y_t=\potY{A}{t}$ almost surely for $t=0,1$
\end{assumption}
\vspace{-0.85cm}
\renewcommand{\theassumption}{1b}
\begin{assumption}
\textit{No Causal Anticipation}: $\potY{1}{0} = \potY{0}{0}$ almost surely. 
\end{assumption}
It is then straightforward to deduce identification of the additive average causal effect of treatment on the treated (ATT) for the
follow-up outcome $\obsY{1}$, i.e., $\psi_{ATT} = E\left(  \potY{1}{1}
-\potY{0}{1}  |A=1\right) =E\left(  \obsY{1}  -\obsY{0}  |A=1\right)  -E\left(
\obsY{1}  -\obsY{0}  |A=0\right)  $, 
justifying DiD.

Despite its popularity, PT has several limitations: (i) it can be violated when dealing with naturally constrained outcomes such as binary or count variables, (ii) it restricts the exposure mechanism and time-varying properties of the outcomes, and (iii) it is scale-dependent; see Section \ref{sec:supp:Limitation PT} of the Appendix for details. In the next Section, we describe an alternative to PT, which accommodates (i)-(iii).

\section{Universal DiD}   \label{sec:UDiD}

\subsection{Identification via Odds Ratio Equi-confounding} \label{sec:UDiD overview}

We introduce a parametrization for a unit's contribution to the likelihood for the potential outcome $\potY{0}{t}  $ conditional on $A$ and observed baseline covariates $X$, assuming independent and identically distributed (i.i.d.) sampling. Let 
\begin{align}
&
h_t(y,x) = f( \potY{0}{t} = y \cond A = 0,X=x)
\label{definition:h}
\\
&
\beta_{t}\left(  y,x\right)  =\log\frac{f\left(  \potY{0}{t}
=y|A=1,X=x\right)  f\left(  \potY{0}{t}=y_{\text{ref}}|A=0,X=x\right)  }{f\left(
\potY{0}{t}=y_{\text{ref}}|A=1,X=x\right)  f\left(  \potY{0}{t}
=y|A=0,X=x\right)  }\ 
\label{definition:odds ratio}
\end{align}
The function $h_{t}\left(  y,x\right)$, referred to as a baseline density, represents the conditional distribution of the potential outcome $\potY{0}{t}$ given $X$ and $A=0$ (i.e., baseline treatment), and the function $\beta_t(y,x)$ is the log of the generalized odds ratio function \citep{Chen2007, TTRR2010} where $\beta_t(y_{\text{ref}},x) = 0$, with $y_{\text{ref}}$ a user-specified reference value; without loss of generality, we take $y_{\text{ref}}=0$. Thus, $\beta_t(y,x)$ encodes the association between $\potY{0}{t}$ and $A,$ evaluated at $y$ given $X=x$. Therefore,
$\beta_{t}\left(  y,x\right)  =0$ for all $y$ encodes no unmeasured confounding given $X=x$, while $\beta_{t}\left(  y,x\right)
\not =0$ quantifies the degree of unmeasured confounding bias at a distributional level. 

These functions parametrize the conditional density of $\potY{0}{t}$ given $(A=a,X=x)$ as $f( \potY{0}{t} = y \cond A = a, X=x) 
\propto 
h_t(y,x)
\exp \big\{ \beta_t(y,x) a \big\}$;  here, $\propto$ stands for the left-handside being proportional to the right-handside for fixed $a$, with proportionality constant equal to the normalizing
constant $\sum_{y}h_{t}\left(  y,x\right)  \exp\left\{  \beta_{t}\left(
y,x\right)  a\right\}  $ $<\infty$ for all $x$, ensuring that the left-handside is a proper
density or probability mass function, and the symbol $\sum_{y}$ may be interpreted as an integral if $y$ is continuous.  This likelihood parametrization can in principle be used to represent any
proper likelihood function one might encounter in practice, i.e., the above formulation is fully unrestricted (or nonparametric) \citep{Chen2007, TTRR2010}. Under consistency and no anticipation, $f\left(  \potY{0}{0}  =y|A=a,X=x\right)  =f\left( \obsY{0}  =y|A=a,X=x\right)$ for $a=0,1$ and $f\left(\potY{0}{1}  =y|A=0,X=x\right)  =f\left(  \obsY{1}=y|A=0,X=x\right)$, respectively, establishing identification of $\beta_0(y,x)$, $h_0(y,x)$, and $h_1(y,x)$ from equations \eqref{definition:h} and \eqref{definition:odds ratio}.  In contrast, identification of  $f\left(  \potY{0}{1}
=y|A=1,X=x\right)  $ and $\beta_{1}\left(
y,x\right)  $ cannot be obtained without an additional condition because $\potY{0}{1}  $ is not observed for units with $A=1$. Our approach relies on the key assumption:
\renewcommand{\theassumption}{2}
\begin{assumption}
Odds Ratio Equi-confounding: $\beta_{0}\left(  y,x\right)  =\beta_{1}\left(  y,x\right)$ for all $(y,x)$.
\end{assumption}
The assumption states that the degree of confounding captured on the log-odds ratio
scale is stable over time, an assumption first considered by \citet{DiD_OREC2022} which they
refer to as odds ratio equi-confounding (OREC). Under OREC, it follows that $f\left(  \potY{0}{1}  =y|A=1,X=x\right)  \propto h_1(y,x)
\exp\left\{  \beta_{0}\left(  y,x\right)  \right\}$, establishing nonparametric identification of $f\left(  \potY{0}{1}  =y|A=1,X=x\right)$ in the sense that $\beta_{0}\left(  y,x\right)$,
$f\left(  \obsY{0}  =y|A=0,X=x\right)  $, and $f\left(  \obsY{1}
=y|A=0,X=x\right)  $ are unrestricted.  Furthermore, one may identify the additive ATT with the expression:%
\begin{align}       \label{eq-OR_representation}
\psi_{ATT}
=E\left(  \obsY{1}  |A=1\right)  -
E \bigg[
\frac{E\left[  \obsY{1}
\exp\left\{  \beta_{0}\left(  \obsY{1},X  \right)  \right\}
|A=0,X\right]  }{E\left[  \exp\left\{  \beta_{0}\left(  \obsY{1},X
\right)  \right\}  |A=0,X\right]  }%
\, \bigg| \, A=1
\bigg]
\end{align}
as we demonstrate in Section \ref{sec:supp:Ident} of the Appendix.

We briefly review examples of special interest where UDiD yields intuitive alternatives to DiD when PT is violated. Suppressing $X$, the ATT satisfies:
\begin{align}
\psi_{ATT}
&
=
\underset{\text{Crude estimand}}{\underbrace{ \big\{ E\left(  \obsY{1}
|A=1\right)  - E\left(  \obsY{1}  |A=0\right) \big\} } }
-
\underbrace{
\big\{ 
E \left( \potY{0}{1}  |A=1 \right)
-
E\left(  \obsY{1}  |A=0\right) 
\big\}
}_{\text{De-biasing Term}}
\ .
\label{general:ATT}
\end{align}
Here, the crude estimand is the difference between the conditional means of the outcome in treated and control groups in the post-treatment period. The de-biasing term is the difference between the ATT and the crude estimand, reflecting unmeasured confounding bias. If there is no unmeasured confounding the bias term vanishes and, consequently, the crude estimand identifies the ATT. Otherwise, suppose that $\potY{0}{t}   |(A=a) \sim N\left(  \mu_{t}\left(  a\right)  ,\sigma_{t}^{2}\right)$. Then, PT and OREC in this model imply that the bias term is equal to $\mu_{1}(1) - \mu_{1}(0) = \mu_{0}(1) - \mu_{0}(0)$ and $\sigma_{1}^{-2} \big\{ \mu_{1}(1) - \mu_{1}(0) \big\} = \sigma_{0}^{-2} \big\{ \mu_{0}(1) - \mu_{0}(0) \big\}$, respectively, and the  corresponding de-biasing terms are :
\begin{align}
&
\text{PT} 
&&
\Rightarrow
&&
E ( \potY{0}{1}  \cond A=1 )
-
E ( \obsY{1} \cond A=0 )
=
\mu_{0}(1) - \mu_{0}(0)
\tag{DiD de-biasing term}
\\
&
\text{OREC} 
&&
\Rightarrow
&& 
E ( \potY{0}{1}  \cond A=1 )
-
E ( \obsY{1} \cond A=0 )
=
\frac{\sigma_{1}^{2}}{\sigma_{0}^{2}}\left\{ \mu_{0}(1) - \mu_{0}(0) \right\}
\tag{UDiD de-biasing term}
\end{align}
Therefore, UDiD reduces to standard DiD if $\sigma_{1}^{2}=\sigma_{0}^{2}$, i.e., the scale of the outcome at $t=0$ matches that at $t=1$; however, UDID is more flexible than DiD in the sense that if $\sigma_{1}^{2}\not =\sigma_{0}^{2},$ UDiD rescales the standard DiD de-biasing term $\mu_{0}\left(  1\right)  -\mu
_{0}\left(  0\right)  $ of the crude estimate to account for a potential
difference of scales between $t=0$ and $t=1$.
Additionally, Section \ref{sec:supp:multiplicative effect} of the Appendix provides analogous comparisons on multiplicative and odds ratio scales.

At this juncture, it is instructive to consider a structural model for OREC analogous to \eqref{eq-standard DiD} which we adopt following \citet{DiD_OREC2022}. Suppressing covariates, consider the following model:
\begin{align}
\label{eq-model OREC}
\tag{UDiD Model}
&
\potY{0}{t} \indep A \cond U_t \ , \ t=0,1, 
\quad \quad
A \cond (U_1=u) \stackrel{D}{=} A \cond (U_0=u) \ , \ \forall u \ ,
\\
\nonumber 
&
U_1 \cond (A=0,Y_1=y) \stackrel{D}{=} U_0 \cond (A=0,Y_0=y) \ , \ \forall y \ .
\end{align}
In \eqref{eq-model OREC}, the relationship between $\potY{0}{t}$ and $U_t$ is unrestricted, while the exposure mechanism of $A$ given $U_t$ is assumed not to depend on time. In addition, \eqref{eq-model OREC} assumes that the conditional distribution of $U_t$ evaluated at $u$ given $(A=0,Y_t)$ is stable over time but otherwise unrestricted. Unlike \eqref{eq-standard DiD}, \eqref{eq-model OREC} is scale-invariant in that any monotone transformation of an outcome that satisfies \eqref{eq-model OREC} remains in the model. Of note, unlike \eqref{eq-standard DiD}, \eqref{eq-model OREC} is agnostic about the presence of an additive interaction between the treatment and $U_t$. In Section \ref{sec:supp:DGP} of the Appendix, we establish that \eqref{eq-model OREC} implies OREC. There, we also describe alternative data generating processes for OREC that may be of independent interest.

For estimation and inference, one may posit GLMs for the outcome process by specifying parametric models for $h_1(y,x)$ and $\beta_{t}(y,x)$. The parameters for these functions can be estimated by standard maximum likelihood theory, which can be easily performed using off-the-shelf software, such as \texttt{geex} package in R \citep{GEEX2020}. As this approach essentially amounts to methods previously described by \citet{Wooldridge2022, Zika2022}, details are relegated to Section \ref{sec:UDID-GLM} of the Appendix. An alternative semiparametric approach that obviates the need to specify a likelihood for the outcome process is given next.

\subsection{Universal DiD Estimation via Extended Propensity Score Weighting }\label{sec:UDID-EPS}

In many real-world applications, GLMs for the outcome can be misspecified, particularly when the distribution of the outcome, such as zero-inflated or truncated outcomes, poses significant challenges for maximum likelihood estimation approaches developed under standard GLM framework. 
To resolve this issue, we provide an alternative approach that uses an extended propensity score model. The approach generalizes the standard propensity score model for the treatment \citep{Rosenbaum1983} to accommodate unmeasured confounding by incorporating the treatment-free potential outcome in the propensity score model. 

The approach is motivated by the invariance property of odds ratios, which provides the following alternative interpretation of $\beta_{t}(y,x)  $ to that given in equation \eqref{definition:odds ratio}, namely
\begin{align*}
& \beta_{t}\left(  y,x\right)  
=
\log
\left\{
\frac{\Pr\left(  A=1| \potY{0}{t}
=y,X=x\right)  \Pr\left(  A=0| \potY{0}{t} =0,X=x \right)  }{ \Pr\left(  A=0|\potY{0}{t}  =y,X=x \right) \Pr\left(
A=1| \potY{0}{t}  =0,X=x \right) } 
\right\}\\
\Leftrightarrow \quad
& 
\log\frac{\pi_{t}\left(  y,x\right)  }{1-\pi_{t}\left(  y,x\right)  }  =\delta_{t}(x)+\beta_{t}\left(  y,x\right)  
\ , \
\delta_{t}(x)
=\log\frac{\Pr\left(  A=1|\potY{0}{t}  =0,X=x\right)
}{\Pr\left(  A=0|\potY{0}{t}  =0,X=x\right)  } \ . 
\end{align*}
where the function $\pi_{t}\left(  y,x\right) =\Pr\left(  A=1|\potY{0}{t} =y,X=x\right)$ is referred to as the \textit{extended propensity score} (EPS) function. 
This alternative obviates the need for specification of a model for $h_{t}\left(  y,x\right)$.  Specifically, under OREC, one can posit a model for the log-odds ratio functions as $\beta_{1}\left(  y,x\right)  =\beta _{0}\left(  y,x\right)  =\alpha_0\T S_{0}(y,x)$ where $S_{0}(y,x)$ is a
user-specified sufficient statistic for the confounding odds ratio parameter. For instance, let $S_{0}(y,x) = (y,yx\T)\T$, i.e., the log-odds of the EPS have a linear relationship in $y$ given $x$, with the corresponding parameter $\alpha_0$. Also, specify a parametric model for $\delta_t(x)$, say $\delta_t(x) =  (1,x\T) \eta_t$, with similar interpretation. Then, $\left(  \eta_0,\alpha_0\right)  $ can be estimated via a standard logistic regression of $A$ on $(1,X,S_{0}\left(
\obsY{0},X \right) )  $. 
Next, $\eta_1$ can be identified as the solution to the population moment equation: 
\begin{align} \label{eq-EPS time 1}
E\left\{ \frac{  1-A   }{1-\pi_{1}\left( \obsY{1}, X
\right)  }\right\}  =1
\Leftrightarrow
E\bigg[  \bigg( \, \begin{matrix}
1 \\[-0.15cm] X
\end{matrix} 
\,
\bigg)
\left(  1-A\right)  \left[
1+\exp\left\{  (1,X\T) \eta_1 +\beta_{0}\left( \obsY{1} ,X \right)  \right\}
\right]  -1 \bigg]  =0 \ .
\end{align}
Therefore,  $\eta_{1}$ can be estimated using the empirical analogue to
the above equation. 
Propensity score weighting UDiD estimation of the ATT based on the estimated EPS follows from the following identifying expression:
\begin{align}  \label{eq-IPW_representation}
\psi_{ATT}
=E\left(  Y_1  |A=1\right)  -\frac{E\left[ \left(  1-A\right)  Y_{1} \exp\left\{ (1,X\T)\eta_1 +  \alpha_0\T S_{0}( Y_1,X )  \right\} \right] }{E\left[ \left(  1-A\right)  \exp\left\{ (1,X\T)\eta_1 +  \alpha_0\T S_{0}( Y_1,X )  \right\} \right]  }
\ .
\end{align}
A proof of this claim is included in Section \ref{sec:supp:Ident} of the Appendix.  


The proposed GLM-based and propensity score weighting UDiD methods rely on (i) modeling the association between covariate and outcome in a GLM, and (ii) modeling the association between covariate and treatment in the EPS model. However, this might introduce a concern that such modeling might introduce specification bias. To address this, we construct a doubly robust \citep{Scharfstein1999, Lunceford2004, Bang2005} estimator for the ATT, in the sense that the estimator is consistent and asymptotically normal if the conditional odds ratio function encoding the association between the treatment-free potential outcome and the treatment is correctly specified conditional on covariates, and either (i) the outcome conditional models for baseline and follow-times, or (ii) the treatment mechanism conditional on the treatment-free potential outcome at its reference value at baseline and follow-up, but not necessarily both, are correctly specified. Details of the doubly robust approach can be found in Section \ref{sec:supp:DR} of the Appendix. Furthermore, to enhance the robustness of the odds ratio function specification particularly for continuous exposure, a flexible yet simple approach might be to posit a model for a discretized version of the outcome, similar to a histogram estimator of a density. In brief, the approach involves converting the outcome into $M$ dummy variables based on the $100(m/M)$th percentiles ($m=1,2,\ldots,M-1$) of the empirical distribution of pre-exposure outcome values. 
Using this discretized outcome, one then can specify the odds ratio function in the UDiD framework; see Section \ref{sec:supp:Diagnosis} of the Appendix for details.

\section{Application: Zika Virus Outbreak in Brazil}            \label{sec:Data}

In 2015, a Zika virus outbreak occurred in Brazil, resulting in more than 200,000 cases by 2016 \citep{Zika2018}. Zika virus infection during pregnancy can affect fetal brain development and lead to severe brain defects such as microcephaly \citep{Zika2016}. Consequently, previous investigators \citep{Zika2018_3, Zika2018_2} conjecture that the Zika virus epidemic might have resulted in a decrease in the birth rate, attributed to individuals' tendency to postpone pregnancy or increase the likelihood of abortion due to the fear of Zika virus-related microcephaly.

We illustrate UDiD with a re-analysis of a study of the effects of an outbreak of Zika virus on birth rate in Brazil originally published in \citet{Zika2022}. Specifically, we consider 2014 and 2016 as pre- and post-exposure periods, respectively, and municipalities in the northeastern state Pernambuco (PE) and those in the southernmost state Rio Grande do Sul (RS) as study units. According to a report from the Brazilian Ministry of Health \citep{Zika2017_Brazil3}, the epidemic was more severe in the northeastern region of Brazil compared to the southern region. In addition, out of 1248 microcephaly cases that occurred in Brazil as of November 28 2015, 646 (51.8\%) cases were reported in PE \citep{Zika2015}, whereas less than 10 cases of Zika-related microcephaly were reported in RS \citep{Zika2017_Brazil2}. Based on the information, individuals in northeastern states may have been more concerned about Zika virus-related microcephaly than those in southern states, which could have contributed to a possible decrease in the birth rate in northeastern states. In contrast, individuals in southern states in Brazil may have experienced minimal behavioral changes compared to those in other regions, as these states were least impacted by the Zika epidemic. Moreover, given the substantial geographical separation of over 2000 kilometers between PE and RS, it is plausible that the behavioral influence stemming from the Zika virus outbreak in PE had limited spillover effects on the population in RS; see \citet{Zika2022} for a related discussion. Therefore, we categorize PE and RS as the treated and control groups, respectively. 

As the pre- and post-exposure outcomes, we use birth rates in 2014 and 2016, respectively, where the birth rate is defined as the total number of live births per 1,000 persons. We treat birth rate as a normally distributed variable in the parametric GLM formulation. 
We focus on 673 municipalities with complete data on the pre- and post-exposure outcomes and treatment, where 185 and 488 municipalities belong to PE and RS, respectively. To further address variation across municipalities due to population differences, we further adjusted for population size, population density, and proportion of females as covariates. The crude mean difference $E(\obsY{1} \cond A=1) - E(\obsY{1} \cond A=0)=3.384$ births per 1,000 persons, indicating that PE region showed higher birth rate than RS region in 2016 despite the Zika virus outbreak.

Based on the proposed approach, we obtain estimates of the additive ATT; see Section \ref{sec:supp:Estimation} of the Appendix for details on the specific steps used for estimation. 
The baseline densities for the outcomes are specified to follow normal distributions  $\potY{0}{t} \cond (A=0,X=x) \sim N( \mu_{t}(x), \sigma_{t}^2 )$ where $\mu_t(x)$ is specified as $\mu_t(x) = (1,x\T) \tau_{t}$. 
Therefore, under an odds ratio parametrization, the odds ratio function is represented as $\beta_0(y,x) = (y,y x\T) \alpha_0$. 
Maximum likelihood estimators of $(\tau_0,\tau_1,\sigma_0^2,\sigma_1^2,\alpha_0)$ are obtained by maximizing the log-likelihood function of the pre-exposure data, i.e., $( Y_0,A,X )$, and the post-exposure data under control, i.e., $( Y_1,A=0,X )$, implemented with \texttt{geex} \citep{GEEX2020}. The EPS is specified as $\pi_{t}(y,x)/\{1-\pi_{t}(y,x)\}= \exp\{ \delta_t(x) + \beta_0(y,x) \}= \exp\{ (1,x\T) \eta_{t} + (y,y x\T) \alpha_0 \}$.
Again, maximum likelihood estimators of $(\eta_0,\alpha_0)$ are obtained by maximizing the log-likelihood function of the pre-exposure data using the same software. We can then estimate $\eta_1$ by solving the empirical analogue of \eqref{eq-EPS time 1} with the estimated odds ratio function using the same software. 
Using these specifications, we obtain six estimates from GLM-based, propensity score weighting, and doubly robust UDiD approaches. We compare these estimates to those derived under PT obtained from \texttt{att\_gt} function implemented in \texttt{did} R package \citep{didpackage}.

Table \ref{Tab-1} summarizes the data analysis results. 
The GLM-based and propensity score weighting UDiD estimates show the largest and smallest effect estimates of $-1.487$ and $-0.831$ births per 1000 persons, respectively. In addition, effect estimates under PT are of a similar value as those obtained from the UDiD approaches, and the corresponding confidence intervals overlap with each other. Compared to the crude estimate of 3.384 births per 1000 persons, the negative effect estimates suggest the presence of substantial confounding bias. This analysis provides compelling evidence that the Zika virus outbreak led to a decline in the birth in Brazil, corroborating similar findings in the literature \citep{Zika2018_3,Zika2018_2,Zika2022}, and further indicating that PT and OREC estimates are of similar magnitude (noting that confidence intervals are wider for UDiD under OREC than for standard DID under PT) and thus, estimates of the magnitudes of effect are not particularly sensitive to the specific nature of equi-confounding assumption used for causal identification.  

\begin{table}[!htp]
\footnotesize
\renewcommand{\arraystretch}{1.1} \centering
\setlength{\tabcolsep}{7pt}
\begin{tabular}{|cc|ccc|}
\hline
\multicolumn{2}{|c|}{\multirow{2}{*}{Estimator}}                                                                                    & \multicolumn{3}{c|}{Statistic}                                                \\ \cline{3-5} 
\multicolumn{2}{|c|}{}                                                                                                              & \multicolumn{1}{c|}{Estimate} & \multicolumn{1}{c|}{SE}    & $95\%$ CI        \\ \hline
\multicolumn{1}{|c|}{\multirow{3}{*}{\begin{tabular}[c]{@{}c@{}}UDiD\\ under OREC\end{tabular}}}       & GLM-based                  & \multicolumn{1}{c|}{-1.487}   & \multicolumn{1}{c|}{0.340} & (-2.153, -0.821) \\ \cline{2-5} 
\multicolumn{1}{|c|}{}                                                                                 & Propensity score weighting & \multicolumn{1}{c|}{-0.831}   & \multicolumn{1}{c|}{0.377} & (-1.570, -0.091) \\ \cline{2-5} 
\multicolumn{1}{|c|}{}                                                                                 & Doubly-robust              & \multicolumn{1}{c|}{-0.974}   & \multicolumn{1}{c|}{0.342} & (-1.645, -0.303) \\ \hline
\multicolumn{1}{|c|}{\multirow{3}{*}{\begin{tabular}[c]{@{}c@{}}Standard DiD\\ under PT\end{tabular}}} & GLM-based                  & \multicolumn{1}{c|}{-1.171}   & \multicolumn{1}{c|}{0.151} & (-1.467, -0.875) \\ \cline{2-5} 
\multicolumn{1}{|c|}{}                                                                                 & Propensity score weighting & \multicolumn{1}{c|}{-1.091}   & \multicolumn{1}{c|}{0.143} & (-1.371, -0.811) \\ \cline{2-5} 
\multicolumn{1}{|c|}{}                                                                                 & Doubly-robust              & \multicolumn{1}{c|}{-1.124}   & \multicolumn{1}{c|}{0.159} & (-1.435, -0.813) \\ \hline
\end{tabular}
\caption{\footnotesize Summary of Data Analysis. Values in ``Estimate'' column represent the additive ATT of the Zika outbreak on the birth rate within the PE region, i.e., the difference between PE's observed average birth rate to a forecast of what it would have been had the Zika outbreak been prevented. Values in ``SE'' and ``95\% CI'' columns represent the standard errors associated with the estimates and the corresponding 95\% confidence intervals, respectively. The reported values are expressed as births per 1,000 persons.}
\label{Tab-1}
\end{table}

In Section \ref{sec:supp:Analysis Discrete} of the Appendix, we provide additional data analysis results for a more flexible discretized odds ratio function. The UDiD estimates using discretized odds ratio are much closer to each other than those reported in Table \ref{Tab-1}, demonstrating less model dependence, with substantially tighter confidence intervals. 

Additional sensitivity analyses inspecting the extent to which empirical findings are sensitive to violation of identifying assumptions are given in Section \ref{sec:supp:Cov Invariance} of the Appendix.

\section{Concluding Remarks}        \label{sec:Conclusion}
In this paper, we have described UDiD as an alternative to standard DiD, which can accommodate outcomes of any type and causal effect estimands possibly defined on nonlinear scales. For UDiD inference, we have described three alternative approaches targeting the average effect of treatment on the treated, the first involves modeling the pre- and post-exposure outcome process, while the second involves positing a model for the EPS, and the third carefully combines both approaches to produce an estimator which possesses a desirable double robustness property of remaining unbiased for the treatment effect, if either outcome model or treatment model is correctly specified. 

While standard DiD which relies for validity on PT, validity of UDiD invokes an assumption of OREC, that the degree of confounding bias encoded with an odds ratio association between the treatment and the treatment-free potential outcome at follow-up is exactly equal to that with the pre-exposure outcome. Realistically, this assumption might not hold exactly in all applications, but can often be expected to be approximately correct if the time between the pre-exposure and post-exposure outcomes is not too large, so that though changing, the magnitude of confounding bias may be expected to evolve smoothly in time at a relatively slow rate. From this perspective, similar to unconfoundedness (a structural assumption that rules out unmeasured confounding), OREC can be logically understood as a structural conditional independence assumption about the distribution of $U_t$ over time while accommodating the potential for unmeasured confounding. This interpretation provides a natural anchoring from which a sensitivity analysis can be initiated. Specifically, one might entertain a sensitivity analysis to the OREC assumption in which the odds ratio function $\beta_{1}\left(  y,x\right)$ is set to $\beta_{1}\left(  y,x\right)=\beta_{0}\left(  y,x\right)+\Delta(y,x)$ where $\Delta(y,x)$ is a user-specified non-identifiable sensitivity function that encodes a potential departure from OREC. One would then proceed by repeating the proposed analyses and reporting various updated estimates of the causal effect of interest over various choices of $\Delta$, thus providing an evaluation of the sensitivity of inferences to possible violations of the OREC condition; see Appendix \ref{sec:supp:sensitivity} for details of such a sensitivity analysis and an application to the Zika study.


Although many DiD methods, including UDiD, are developed assuming no interference \citep{Cox1958}, it is important to acknowledge that interference may be plausible in various applications, particularly in the context of infectious diseases. For example, in our data analysis, interference could have occurred if the Zika virus outbreak in PE had resulted in significant changes in behavior among individuals in RS. Hence, in order to properly use DiD methods, it is crucial in practice to assess the possibility of interference in the context in view. In the event that interference becomes a concern, one might consider DiD methods that are explicitly designed to account for interference \citep{DiD_Spillover_2017, DiD_Spillover_2021}. In addition, given its relevance in epidemiological applications, this paper has mainly focused on the additive ATT, and the comparison between OREC and PT. Nonetheless, OREC has the capacity to accommodate nonlinear treatment effects, such as quantile treatment effects on the treated, as long as the treatment effect in view is uniquely defined as the solution to a moment equation. As a result, comparing OREC and identifying assumptions for nonlinear treatment effects in DiD settings could provide useful insights. We refer interested readers to \citet{DiD_OREC2022} for details.

An important potential generalization of UDiD concerns settings where richer longitudinal data might be available for each unit. In such settings, one might be able to leverage past outcomes to either validate or relax OREC; a possibility we plan to explore in the future. In addition, in panel data settings, staggered treatment initiation might occur, in which case various generalizations of OREC might be possible, thus effectively extending recent developments under PT to handle such complex study designs \citep{Callaway2021,Dukes2022_BespokeIV,Shahn2022}. 
\newpage

\appendix

\section*{Appendix}

\section{Details of the Main Paper}

\subsection{Limitation of Parallel Trends} \label{sec:supp:Limitation PT}

The parallel trends (PT) assumption is generally not subject to an empirical test as it does not impose any restriction on the observed data distribution. A notable partial exception is that of bounded outcomes, in which case PT may be refuted if it empirically implies potential outcome predictions that fall outside of the domain of the outcome. As a concrete example, suppose that a binary treatment-free potential outcome is distributed as $ \potY{0}{0} \cond A \sim \text{Ber}(0.4 + 0.4A)$ and $\potY{0}{1} \cond A \sim \text{Ber} (0.8 + 0.16A)$. Under PT, the conditional mean $\EXP \big( \potY{0}{1} \cond A=1 \big)$ is evaluated as $\EXP \big(Y_1 \cond A=0 \big) + \EXP \big(Y_0 \cond A=1 \big) - \EXP \big(Y_0 \cond A=0 \big) = 1.2$. It is different from its true value is 0.96 as well as falling beyond its natural range given by the unit interval. As a result, PT may not always be credible in the case of naturally bounded outcomes. 

Moreover, imposing PT involves a trade-off between restrictions on the exposure mechanism and restrictions on the time-varying properties of the outcomes \citep{Ghanem2022}, a compromise that may not always be appropriate in health applications. Another limitation of PT is that it is scale-dependent, i.e., given an outcome that satisfies the PT assumption, a monotone transformation of the outcome might not in general satisfy the assumption without an extra condition beyond PT \citep{RothSantAnna2023}. For instance, we focus on \eqref{eq-standard DiD}, which is restated below:
\begin{align}
    &
\potY{0}{t} = h(U_t,t)
\ , \ 
\quad \quad
h(u,t) = u + \beta_T t
\ , \
\quad \quad
U_t = \beta_0 + \beta_A A + \epsilon_t 
\ , \
\tag{DiD Model}
\\
\nonumber
&
\epsilon_t \text{ satisfies either }
\Bigg\{
\begin{array}{ll}
\text{time independence: } 
& \epsilon_1 | A \stackrel{D}{=} \epsilon_0 | A
\quad \text{ or }
\\
\text{treatment independence: } 
&
\epsilon_t | (A=0) \stackrel{D}{=} \epsilon_t | (A=1) 
\end{array}
\ .
\end{align}
Then, PT is generally not satisfied if $h(u,t)$ is nonlinear in $u$, say $h(u,t) = \phi(u+\beta_T t)$ for a nonlinear monotone transformation $\phi (\cdot)$.

\subsection{Universal DiD Estimation using Generalized Linear Models}  \label{sec:UDID-GLM}
For estimation and inference in practice, it is convenient to consider a familiar and widely used class of parametric
models compatible with the above likelihood formulation which consists of the
exponential family, obtained by specifying $\beta_{t}\left(  y,x\right)
=\alpha_{t}\T S_{t}(y,x)$, where $\alpha_{t}$ is an unknown vector of
canonical parameter and $S_{t}(y,x)$ is a corresponding vector of sufficient statistics. Many common distributions are included in this class of parametric
models, including Logistic regression for dichotomous
outcomes, Poisson regression for count outcomes, and linear regression for
Gaussian outcomes. Under no covariate, the first two admit a sufficient statistic $S_{t}(y)=y,$ while the Gaussian model admits the sufficient statistic $S_{t}(y)=(y,y^{2}%
)\T$. The baseline densities $h_{t}\left(  y\right)  $ of these three
models are Bernoulli, Poisson and Gaussian, respectively. 
As a concrete example, suppose that the outcome is binary and is distributed as $\potY{0}{t} \cond (A=a) \sim \text{Ber} (p_t(a)) $. The baseline densities are given by  $h_t(y) = \{ p_t(0) \}^y \{ 1-p_t(0) \}^{1-y}$ and the log-odds ratio functions are of the specific form $\beta_t(y) = \alpha_t S_t(y)$ where $S_t(y)=y$ is a sufficient statistic, and $\alpha_t$ is the log-odds ratio relating $\potY{0}{t}$ and $A$ at time $t$, i.e., 
\begin{align*}
\alpha_t = \log
\left[ \frac{p_t(1) \{ 1- p_t(0) \} }{ \{ 1- p_t(1) \} p_t(0) } \right] \ .
\end{align*}
Therefore, OREC encodes that the odds ratios relating $A$ and $Y^{a=0}_{t}$ at time $t=0$ and $t=1$ are equal. 
When covariates are available, one may specify $S_t(y,x) = (y,y x\T)\T$. Beyond these examples, many other common parametric models are available within the exponential family, and an exhaustive list is provided in the next paragraph, with their corresponding sufficient statistics. Estimation and inference under such
parametric models follow immediately from standard maximum likelihood theory
and is easily performed using off-the-shelf software, such as \texttt{geex} package in R \citep{GEEX2020}. Thus, under such parametric specification of the likelihood, OREC delivers identification of the counterfactual density
\[
f\left(  \potY{0}{1}  =y|A=1, X=x\right)  \propto h_{1}\left(
y,x \right)  \exp\left\{  \alpha_{0}\T S_{0}(y,x) \right\} 
\]
which can then be used as basis for UDiD inferences.  

Next, we provide a comprehensive collection of exponential family distributions with corresponding sufficient statistics. 

\begin{itemize}
\item[(1)] Gaussian: $\potY{0}{t} \cond A=a \sim N \big( \mu_t(a), \sigma_t^2 (a) \big) $ with support $y \in (-\infty , \infty)$
\begin{align*}
& 
h_{t}(y)
=
\big\{ 2 \pi \sigma_t^2 (0) \big\}^{-1/2}
\exp \bigg[
- \frac{ \big\{ y - \mu_t(0) \big\}^2 }{2\sigma_t^2 (0) }
\bigg] \  , 
\\
& \alpha_t 
=
\Bigg(
\begin{array}{c}
\mu_t(1)/\sigma_{t}^{2}(1) - \mu_t(0)/\sigma_{t}^{2}(0) 
\\
- 1/ \{ 2 \sigma_{t}^{2}(1)\} + 1/\{2 \sigma_{t}^{2}(0) \}
\end{array}
\Bigg)
\ , \ 
S_{t}(y)
=
\Bigg(
\begin{array}{c}
y
\\
y^2
\end{array}
\Bigg)
\end{align*}
We remark that if the variance does not depend on treatment status, i.e., $\sigma_{t}^2 = \sigma_{t}^2(0)=\sigma_{t}^2(1)$, $\alpha_t$ and $S_{t}(y)$ are reduced to $\alpha_t = \sigma_{t}^{-2} \{ \mu_t(1)  - \mu_t(0) \} $ and $S_{t}(y)=y$.

\item[(2)] Binomial: $\potY{0}{t} \cond A=a \sim \text{Binomial}\big( M , p_t(a) \big)$ with support $y \in \{0,1,\ldots,M\}$
\begin{align*}
& 
h_t(y) = 
{M \choose y}
\exp \Big[
y \log \Big\{ \frac{p_t(0)}{1-p_t(0)} \Big\} 
+ M \log \big\{ 1 - p_t(0) \big\}
\Big] \ , 
\\
& \alpha_t = \log \Big\{ \frac{p_t(1)}{1-p_t(1)} \Big\} - \log \Big\{ \frac{p_t(0)}{1-p_t(0)} \Big\}
\ , \
S_{t}(y)
=
y
\end{align*}        
We remark that the Bernoulli distribution is a special case when $M=1$.

\item[(3)] Poisson: $\potY{0}{t} \cond A=a \sim \text{Poisson}\big( \lambda_t(a) \big)$ with support $y \in \{0,1,2,3,\ldots \}$
\begin{align*}
& h_t(y) =
(y!)^{-1}
\exp \Big[
- \lambda_t(0) +
y \log \big\{ \lambda_t(0) \big\}
\Big] \ , 
\\
&  \alpha_t = \log \big\{ \lambda_t(1) \big\} - \log \big\{ \lambda_t(0) \big\}
\ , \
S_t(y) = y
\end{align*}

\item[(4)] Negative binomial: $\potY{0}{t} \cond A=a \sim \text{NegBinom} \big(M, p_t(a) \big)$ with support $y \in \{0,1,\ldots\}$
\begin{align*}
& h_t(y) = { y + M  - 1 \choose y } 
\exp \Big[ 
M \log \big\{ p_t(a) \big\} + y \log \big\{ 1- p_t(a) \big\}
\Big]
\\
&
\alpha_t = \log \bigg\{ \frac{1-p_t(1)}{1-p_t(0)} \bigg\} 
\ , \
S_t(y) = y
\end{align*}
We remark that the geometric distribution is a special case when $M=1$.

\item[(5)] Gamma: $\potY{0}{t} \cond A=a \sim \text{Gamma}\big( \kappa_t(a) , \lambda_t(a) \big) $ with support $y \in (0,\infty)$
\begin{align*}
&
h_t(y)
= 
\frac{1}{ \Gamma\big( \kappa_t(0) \big) \lambda_t(a)^{\kappa_t(a)} }
\exp 
\bigg\{
\kappa_t(0) \cdot \log(y)
- \log(y)
+
\frac{y}{\lambda_t(0)}
\bigg\} \ , 
\\
&
\alpha_t 
= 
\Bigg(
\begin{array}{c}
- \{ \lambda_t(1) \}^{-1} + \{ \lambda_t(0) \}^{-1}
\\
\kappa_t(1) - \kappa_t(0)
\end{array}
\Bigg)
\ , \
S_t(y)
=
\Bigg(
\begin{array}{c}
y
\\
\log y
\end{array}
\Bigg)
\end{align*}
We remark that the exponential distribution is a special case of the gamma distribution when $\kappa_t(0)=\kappa_t(1)=1$. In this case, $\alpha_t$ and $S_t(y)$ are reduced to $\alpha_t = - \{ \lambda_t(1) \}^{-1} + \{ \lambda_t(0) \}^{-1}$ and $S_t(y)=y$.

\item[(6)] Pareto: $\potY{0}{t} \cond A=a \sim \text{Pareto} \big( y_m, \kappa_t(a) \big)$ with support $y \in (y_m, \infty)$
\begin{align*}
&
h_t(y) 
=
\kappa_t(0) y_m^{\kappa_t(0)}
\exp 
\Big[
- \big\{ \kappa_t(0) + 1 \big\} \log y
\Big] 
\\
&
\alpha_t
= - \log \kappa_t(1) + \log \kappa_t(0)
\ , \
S_t(y) = \log y
\end{align*}

\item[(7)] Weibull: $\potY{0}{t} \cond A=a \sim \text{Weibull} \big( \kappa_t , \lambda_t(0) \big)$ with support $y \in (0,\infty)$
\begin{align*}
& 
h_t(y) 
= \frac{\kappa_t}{\lambda_t(0)^{\kappa_t}}
\exp
\Bigg[
\big\{ \kappa_t - 1 \big\} \log y
- \bigg\{ \frac{y}{\lambda_t(0)} \bigg\}^{\kappa_t}
\Bigg]
\\
&
\alpha_t = - \big\{ \lambda_t(1) \big\}^{\kappa_t} + \big\{ \lambda_t(0) \big\}^{\kappa_t}
\ , \
S_t(y)
=
y^{\kappa_t}
\end{align*}
We remark that the exponential distribution is a special case when $\kappa_t=1$.

\item[(8)] Laplace: $\potY{0}{t} \cond A=a \sim \text{Laplace} \big( \mu_t , \sigma_{t}(a) \big) $ with support $y \in (-\infty,\infty)$
\begin{align*}
&
h_t(y)
=
\frac{1}{2 \sigma_{t}(0)}
\exp \bigg\{
- \frac{| y - \mu_t |}{\sigma_t(0)}
\bigg\}
\ , 
\\
&
\alpha_t = - \sigma_t^{-1}(1) + \sigma_t^{-1}(0)
\ , \
S_t(y) = \big| y - \mu_t \big|
\end{align*}

\item[(9)] Beta: $\potY{0}{t} \cond A=a \sim \text{Beta} \big( \kappa_t(a), \lambda_t(a) \big)$ with support $y \in (0,1)$
\begin{align*}
&
h_t(y)
=
\frac{ \Gamma\big( \kappa_t(0)+\lambda_t(0) \big) }{\Gamma \big( \kappa_t(0) \big) \Gamma \big( \lambda_t(0) \big)}
\exp \Big\{
\kappa_t(0) \log y + \lambda_t(0) \log (1-y) - \log y - \log (1-y)
\Big\}
\ , 
\\
& \alpha_t 
=
\Bigg(
\begin{array}{c}
\kappa_t(1) - \kappa_t(0)
\\
\lambda_t(1) - \lambda_t(0)
\end{array}
\Bigg) 
\ , \
S_t(y)
=
\Bigg(
\begin{array}{c}
\log y
\\
\log (1-y)
\end{array}
\Bigg) 
\end{align*}
\end{itemize}

\subsection{Multiplicative Causal Effects in Poisson and Binary Outcomes}       \label{sec:supp:multiplicative effect}

In the next two examples, we consider multiplicative causal effects as natural scales to measure causal effects for Poisson and binomial outcomes, respectively.  
Suppose that $\potY{0}{t} $ is a count variable which
follows the distribution $\potY{0}{t}  |A=a\sim$Poisson($\lambda
_{t}(a))$ with mean parameter $\lambda_{t}(a)>0;$ then UDiD identifies the average
causal effect on the treated on the\  multiplicative scale $E\left( \potY{1}{1} |A=1\right)  / 
E\left( \potY{0}{1} |A=1\right)  $  with
\begin{align*}
& 
\frac{ E\left( \potY{1}{1} |A=1\right) }{
E\left( \potY{0}{1} |A=1\right) }
 =\underset{\text{Crude estimand }}{
\underbrace{\frac{E\left(  \obsY{1}  |A=1\right)  }{E\left(  \obsY{1}  |A=0\right)  }}}%
\times\underset{\text{UDiD debiasing term}}{\underbrace{\frac{\lambda_{0}%
(0)}{\lambda_{0}(1)}}}%
\ .
\end{align*}
Therefore, providing justification for debiasing the crude ratio estimand at
$t=1$ with the ratio estimand of the association of the treatment with the
outcome at $t=0.$ The additive version of the causal effect can likewise be deduced to equal $E\left(  \obsY{1}  |A=1\right)  -E\left(  \obsY{1}  |A=0\right)  \lambda_{0}(1)/\lambda_{0}(0).$

Finally, suppose that $\potY{0}{t} $ is a binary variable which
follows the distribution $\potY{0}{t}  |A=a\sim$Bernoulli($p_{t}%
(a))$ with event probability $0<p_{t}(a)<1;$ then UDiD identifies the average
causal effect on the treated on the odds ratio scale with
\begin{align*}
& 
\frac{\Pr\left(  \potY{1}{1}  =1|A=1\right)  \Pr\left(
\potY{0}{1}  =0|A=1\right)  }{\Pr\left(  \potY{1}{1}
=0|A=1\right)  \Pr\left(  \potY{0}{1}  =1|A=1\right)  }
\\
& =\underset{\text{Crude estimand }}{\underbrace{\frac{\Pr\left(  \obsY{1} 
=1|A=1\right)  \Pr\left(  \obsY{1}  =0|A=0\right)  }%
{\Pr\left(  \obsY{1}  =0|A=1\right)  \Pr\left(  \obsY{1}
=1|A=0\right)  }}}\times\underset{\text{UDiD debiasing term}}{\underbrace{\frac
{p_{0}(0)\left\{  1-p_{0}(1)\right\}  }{p_{0}(1)\left\{  1-p_{0}(0)\right\}  } }%
} \ ;
\end{align*}
therefore providing justification for debiasing the crude odds ratio at
\thinspace$t=1$ by the odds ratio association of the treatment with the
outcome at $t=0$. 
These three examples demonstrate the ability of UDiD to naturally adapt to the
scale of the causal estimand in view in order to debias the corresponding
crude estimand.

\subsection{Doubly Robust Universal DiD}   \label{sec:supp:DR}

We have thus far described two separate UDiD estimation strategies under the
OREC assumption; (i) a generalized linear model (GLM) maximum likelihood approach, which relies on correct specification of a model for the outcome, and (ii) an extended propensity score weighted approach, which instead relies on correct specification of a model for the treatment mechanism. Thus, misspecification of either model will likely result in biased inferences.   As in most
practical situations one cannot be sure that either model is correctly specified, the most one can hope for is to obtain unbiased inferences about a causal of interest if either model is correct, without a priori knowledge of which model is incorrect. An estimator with such a property is said to be doubly robust \citep{Scharfstein1999, Lunceford2004, Bang2005}.  Such doubly robust estimators have previously been proposed under unconfoundedness conditions \citep{Bang2005, Tsiatis2006, TTRR2010}, and more recently for DiD methods under standard PT conditions \citep{SantAnna2020, Callaway2021}. Next, we propose a doubly robust UDiD estimator of the effect of treatment on the treated, which essentially amounts to obtaining a doubly robust estimator of the average exposure-free counterfactual outcome in the treated parameter $E\left( \potY{0}{1} |A=1\right)$. A key challenge to constructing a doubly robust estimator in the current setting which does not arise either under unconfoundedness or under PT is that, as previously mentioned, the log-odds ratio model, say 
\begin{equation} 
\beta\left(  y,x\right)  =\alpha_{0}\T S_{0}(y,x);\label{OR}%
\end{equation}
is a parameter shared by both models (i)\ and (ii). Thus, double robustness
necessarily requires correct specification of the log-odds ratio model
$\left(  \ref{OR}\right)  $ and obtaining a doubly robust estimator for the
latter which is consistent if either (i)\ or (ii)\ holds. \ Fortunately, such
a doubly robust estimator was obtained by \citet{TTRR2010},
and is best described by
noting that the following random variable:
\begin{align}       \label{eq-OddsRatioMomentEquation}
\big\{  A-\pi_{0}^{\dagger}\left(  0,X\right)  \big\}  \exp \big\{
-\alpha_{0}\T S_{0}(\obsY{0}  ,X)A \big\}
\left\{
S_{0}(\obsY{0}  ,X)-E^{\dagger}\left(  S_{0}(\obsY{0}
,X)|A=0,X\right)  \right\}    
\end{align}
has conditional mean zero given $X$ provided that (i) $\left(  \ref{OR}\right)  $
is correctly specified, and (ii) either 
$\pi_{0}^{\dagger}\left(  0,X\right)
=\pi_{0}\left(  0,X\right)  $, the true extended propensity score evaluated
at $\obsY{0}=0$, or $E^{\dagger}\left\{  S_{0}(\obsY{0}  ,X)|A=0,X\right\}
=E\left\{  S_{0}(\obsY{0}  ,X)|A=0,X\right\}  $, the true conditional
mean of the sufficient statistic $S_{0}$ with respect to the distribution of
$\obsY{0}  $ given $A=0$ and $X$, but not necessarily both, is correctly specified. 
Throughout, $^{\dagger}$ indicates the
corresponding quantity may or may not match the true data generating
mechanism.  This property motivates an empirical moment equation for
$\alpha_{0}$ which can be solved upon replacing $\pi_{0}^{\dagger}\left(
0,X\right)  $ and $E^{\dagger}\left\{ S_{0}(\obsY{0}
,X)|A=0,X\right\}  $ with corresponding estimator previously obtained. Next, following \citet{Scharfstein1999} and \citet{Liu2020}, 
a doubly robust estimator of $E\left(
\potY{0}{1}  |A=1\right)  $ can be obtained by empirically
evaluating the following equation:
\begin{align}           \label{eq-ATT0-DR}
& E\left[  \frac{1-A  }{\Pr\left(  A=1\right)  }\frac{\pi
_{1}^{\dagger}\left(  \obsY{1}  ,X\right)  }{ 1-\pi_{1}^{\dagger
}\left( \obsY{1} ,X \right)  }
\left[  \obsY{1}
-\frac{E^{\dagger}\left[  \obsY{1}  \exp\left\{  \alpha_{0}^{\prime
}S_{0}( \obsY{1} , X  )\right\}  |A=0,X\right]  }{E^{\dagger}
\left[
\exp\left\{  \alpha_{0}\T S_{0}( \obsY{1} , X  )\right\}
|A=0,X\right]  }\right]  \right.  \nonumber
\\
& 
\hspace*{2cm}
\left.  + {\frac{A}{\Pr(A=1)}} \frac{E^{\dagger}\left[ \obsY{1}  \exp\left\{  \alpha
_{0}\T S_{0}( \obsY{1} , X  )\right\}  |A=0,X\right]  }{E^{\dagger
}\left[  \exp\left\{  \alpha_{0}\T S_{0}( \obsY{1} , X  )\right\}
|A=0,X\right]  }\right]
\end{align}
which as shown in Section \ref{sec:supp:Ident} of the Appendix, identifies $E\left( \potY{0}{1} |A=1\right)  $ if
either $\pi_{1}^{\dagger} (0,X) =\pi_{1}(0,X)$ or
\begin{align*}
\frac{E^{\dagger}\left[  \obsY{1}  \exp\left\{  \alpha
_{0}\T S_{0}( \obsY{1} , X  )\right\}  |A=0,X\right]  }{E^{\dagger
}\left[  \exp\left\{  \alpha_{0}\T S_{0}(\obsY{1} , X  )\right\}
|A=0,X\right] }=\frac{E\left[ \obsY{1}  \exp\left\{  \alpha
_{0}\T S_{0}( \obsY{1} , X  )\right\}  |A=0,X\right]  }
{E\left[  \exp\left\{  \alpha_{0}\T S_{0}(\obsY{1} , X  )\right\}
|A=0,X\right]}. \end{align*} 
A detailed description of the corresponding estimator is provided in the next Section, together with methods for obtaining standard errors and corresponding confidence intervals. 


\subsection{Estimation of the Average Causal Effect of Treatment on the Treated}        \label{sec:supp:Estimation}

For notational brevity, let $\psi_1^\true = E ( \potY{1}{1} \cond A=1 )$ and $ \psi_0^\true = E ( \potY{0}{1} \cond A=1 )$. Let $\Delta(\psi_0,\psi_1)$ be a function that contrasts $\psi_1$ and $\psi_0$. Two popular choices for the contrasting function are the difference $\Delta_{\text{Add}}( \psi_0,\psi_1) = \psi_1-\psi_0$ and the ratio $\Delta_{\text{Mult}}(\psi_0,\psi_1) = \psi_1/\psi_0$. Using these contrasting functions, the additive and multiplicative additive average causal effects of treatment on the treated are represented as $\psi_{\text{Add}}^\true := \Delta(\psi_0^\true,\psi_1^\true)$ and $\psi_{\text{Mult}}^\true := \Delta(\psi_0^\true,\psi_1^\true)$, respectively. However, the approach below is applied to general causal estimands of the form $\Delta(\psi_0^\true,\psi_1^\true)$.

Before we present details, we introduce additional notations. Let $N$ be the number of observed units, indexed by subscript $i = 1, \ldots N$. 
Let $\AVER (g) $ denote the average of function $g(Y_{1},Y_{0},A,X)$ across $N$ units, i.e., $\AVER(g) = N^{-1} \sum_{i=1}^{N} g( Y_{1,i}, Y_{0,i}, A_i, X_i)$. 
Let $\text{expit}(x) = \exp(x)/\{ 1+ \exp(x) \}$.
For random variables $V$ and $W$, let $V \stackrel{P}{\rightarrow} W$ and $V \stackrel{D}{\rightarrow} W$ denote $V$ converges to $W$ in probability and in distribution, respectively.

\subsubsection*{Estimating Equation for Parameters of pre-exposure Models}

Recall that counterfactual version of our likelihood
model can be defined as:
\begin{equation}        \label{eq-outcomelikelihood}
f\left(  \potY{0}{t} =y|A=a,X=x\right)  \propto h_{t}\left(y,x\right)  \exp\left\{  \beta_{t}\left(  y,x\right)  a\right\}
\end{equation}
where $\beta_t(y,x)$ is parametrized as $\beta_t(y,x) = \alpha_\OR \T S_0(y,x)$ for $t=0,1$ under OREC. 
We further assume that baseline density $h_{t}\left(y,x\right)$ belongs to the exponential family
\begin{align}      \label{eq-outcomelikelihood-baseline}
h_{t}\left( y,x \con \tau_t, \gamma_t \right)  \propto h^{\ast}\left(  y ;\tau_{t}\right)
\exp\left\{  \gamma_{t}\T S_{t}^{\ast}\left(  y,x\right)  \right\}
\end{align}
where $h^{\ast}\left(  y ;\tau_{t}\right)  $ is a user-specified distribution
for the baseline density $f\left( \potY{0}{t}   =y|A=0,X=0\right)
$ with unknown parameter $\tau_{t}$, and $S_{t}^{\ast}\left(  y,x\right)  $ is
a user-specified sufficient statistic for $\gamma_{t},$ with  $S^{\ast}\left(
0,x\right)  =0;$ e.g., $S^{\ast}(y,x)=\left(  y,yx\T \right)  \T .$

We start with constructing the estimating equation for the parameters of the pre-exposure outcome regression model. From equations \eqref{eq-outcomelikelihood} and \eqref{eq-outcomelikelihood-baseline}, the log-likelihood of the conditional distribution of $\obsY{0} \cond (A,X)$ is represented as
\begin{align*}
&
\ell_{t=0} ( \obsY{0}, A, X \con \tau_{0}, \gamma_{0} , \alpha_\OR ) 
\\
& 
=
\log h_0^{\ast} (\obsY{0} \con \tau_{0} ) + \gamma_0\T  S_0^{\ast}( \obsY{0}, X ) +  A  \cdot \{ \alpha_\OR\T S_0(y,x) \} + g_0 (\obsY{0},A,X \con \tau_0,\gamma_0,\alpha_{\OR})
\end{align*}
where $g_0$ is a function that makes both hand sides of equation \eqref{eq-outcomelikelihood} equivalent. Under standard regularity conditions on the likelihood function, the maximum likelihood estimators of $(\tau_0,\gamma_0,\alpha_\OR)$ are represented as the solution to the following estimating equation:
\begin{align*}
& 
0
=
\AVER
\Big\{
\Psi_{t=0,\OR} (\obsY{0},A,X \con \widehat{\tau}_{0}, \widehat{\gamma}_{0}, \widehat{\alpha}_\OR)
\Big\}
\ , \\
&
\Psi_{t=0,\OR} (\obsY{0},A,X \con {\tau}_{0}, {\gamma}_{0}, {\alpha}_\OR) 
= \frac{  \partial \ell_{t=0} (\obsY{0},A,X \con \tau_{0},\gamma_{0},\alpha_\OR)  }{ \partial ( \tau_{0},\gamma_{0},\alpha_\OR ) } \ .
\end{align*}
We remark that $\widehat{\alpha}_\OR$ is a preliminary estimator of the odds ratio parameter $\alpha_0$ (see \eqref{OR}), and we will use another estimator of $\alpha$ later.

Next, we construct the estimating equation for the parameters of the pre-exposure extended propensity score model. Following the specification in the main paper, 
one might specify a logistic regression of the form
\begin{align}       \label{eq-propensitylikelihood}
\log\frac{\pi_{t}\left(  y,x\right)  }{1-\pi_{t}\left(  y,x\right)  }=\left(
1,x\T \right)  \eta_t +\alpha_\PS\T S_{0}(y,x) \ .
\end{align}
Based on \eqref{eq-propensitylikelihood}, the maximum likelihood estimators of the parameters $\eta_0$ and $\alpha$ are obtained as the solution to the following estimating equation:
\begin{align*}
& 0 
=
\AVER
\Big\{
\Psi_{t=0,\text{PS}}(\obsY{0},A,X \con \widehat{\eta}_0,\widehat{\alpha}_\PS )
\Big\} \ , \\
&
\Psi_{t=0,\text{PS}}(\obsY{0},A,X \con \eta_0,\alpha_\PS )
=
\begin{pmatrix}
1 \\ X \\ S_0(\obsY{0},X)
\end{pmatrix}
\left[
A - \text{expit} \Big\{ (1,X\T ) \eta_0 + \alpha_\PS\T S_0(\obsY{0},X)  \Big\}
\right] \ .
\end{align*}
We remark that $\widehat{\alpha}_\PS$ is a preliminary estimator of the odds ratio parameter $\alpha_0$ (see \eqref{OR}), and we will use another estimator of $\alpha$ later.


\subsubsection*{Estimating Equation for the Odds Ratio Parameter}

To obtain the estimator of $\alpha_0$, we use the moment equation \eqref{eq-OddsRatioMomentEquation}. 
We first obtain representations of $\pi_0^{\dagger}(0,X)$. 
From the parametric model specified in \eqref{eq-propensitylikelihood}, $\pi_0^{\dagger}(0,X)$ is represented as $\pi_0^{\dagger}(0,X) = \text{expit} \big\{ (1,X\T ) \eta_0 \big\}$. 
Additionally, under the parametric model in \eqref{eq-outcomelikelihood-baseline}, the baseline conditional density of $\obsY{0}$ given $(A=0,X)$ is proportional to $h_0(y , x \con \tau_{0},\gamma_{0})$. Therefore, $E^{\dagger}\{ S_0(\obsY{0},X) \cond A=0,X \}$ can be parametrized by $(\tau_0,\gamma_0)$, i.e., $E^{\dagger}\{S_0(\obsY{0},X) \cond A=0,X \con \tau_0,\gamma_0\}$. For instance, if the baseline outcome is modeled as $\obsY{0} \cond (A=0,X) \sim N(\mu_0(A,X \con \tau_0, \gamma_0) ,\sigma_0^2(X \con \tau_0, \gamma_0))$ and the sufficient statistic is chosen as $S_0(\obsY{0},X)=(1,X\T ) \obsY{0}$, we obtain $E^{\dagger}\{ S_0(\obsY{0},X) \cond A=0,X \con \tau_0,\gamma_0 \} = \mu_0(0,X \con \tau_0, \gamma_0) (1,X\T )\T$ from straightforward calculation.
Using these parametric specifications, the moment equation \eqref{eq-OddsRatioMomentEquation} is represented as follows:
\begin{align*}
& 
\Psi_{\text{Odds Ratio}} 
(\obsY{0},A,X \con \eta_0, \tau_0, \gamma_0, \alpha_0)
\\
&
:=
\big\{  A-\pi_{0}^{\dagger}\left(  0,X \con \eta_0 \right)  \big\} 
\exp\big\{
-\alpha_{0}\T S_{0}(\obsY{0}  ,X)A \big\}
\left[
S_{0}(\obsY{0}  ,X)-E^{\dagger}\left\{  S_{0}(\obsY{0} ,X)|A=0,X \con \tau_0, \gamma_0 \right\}  \right]
\end{align*}
We can obtain the estimator of $\alpha_0$ from the solution to the estimating equation 
\begin{align}           \label{eq-EE_OddsRatio}
0 = \AVER \big\{ \Psi_{\text{Odds Ratio}} (\obsY{0},A,X \con \widehat{\eta}_0, \widehat{\tau}_0, \widehat{\gamma}_0, \widehat{\alpha}_0) \big\}
\end{align}
where $(\widehat{\tau}_{0}, \widehat{\gamma}_{0},\widehat{\eta}_0)$ are obtained from pre-exposure estimating equations. 

\subsubsection*{Estimating Equation for Parameters of post-exposure Models}

From equation \eqref{eq-outcomelikelihood-baseline}, the log-likelihood of the conditional distribution of $\obsY{1} \cond (A=0,X)$ is represented as
\begin{align*}
\ell_{t=1} ( \obsY{1}, A=0, X \con \tau_{1}, \gamma_{1} , \alpha_\OR ) 
=
\log h_1^{\ast} (\obsY{1} \con \tau_{1} ) + \gamma_1\T  S_1^{\ast}( \obsY{1}, X )  + g_1(\obsY{0},A=0,X \con \tau_1,\gamma_1,\alpha_{\OR})
\end{align*}
where $g_1$ is a function that makes both hand sides of equation \eqref{eq-outcomelikelihood-baseline} equivalent. Again, the maximum likelihood estimators of $(\tau_1,\gamma_1)$ are represented as the solution to the following estimating equation: 
\begin{align*}
& 
0
=
\AVER
\Big\{
\Psi_{t=1,\OR} (\obsY{1},A,X \con \widehat{\tau}_{1}, \widehat{\gamma}_{1})
\Big\}
\ , \\
&
\Psi_{t=1,\OR} (\obsY{1},A,X \con {\tau}_{1}, {\gamma}_{1}) 
= \ind(A=0) \cdot \frac{  \partial \ell_{t=1} (\obsY{1},A,X \con \tau_{1},\gamma_{1})  }{ \partial ( \tau_{1},\gamma_{1} ) } \ .
\end{align*}

As discussed in the main paper, $\eta_{1}$ is estimated from the following estimating equation:
\begin{align*}
& 0
=
\AVER \Big\{ \Psi_{t=1,\PS} (\obsY{1},A,X \con \widehat{\eta}_{1},\widehat{\alpha}_{0} ) \Big\} \ ,
\\
&
\Psi_{t=1,\PS} (\obsY{1},A,X \con \eta_{1},\alpha_{0} )
= 
  \left(
\begin{array}
[c]{c}%
1\\
X
\end{array}
\right)  \Big[ \left(  1-A\right)  \left[  1+\exp\left\{  \left(
1,X\T \right)  \eta_{1}+\alpha_0\T S_{0}(\obsY{1},X)\right\}  \right]
-1\Big] 
\end{align*}
where $\widehat{\alpha}_0$ is obtained from \eqref{eq-EE_OddsRatio}. 

\subsubsection*{Estimating Equation for the Treatment Effects}

Since $\psi_1^\true = E(\obsY{1} \cond A=1)$, a natural estimator for $\psi_1^\true$ is the solution to $0 = \AVER \big\{  \Psi_{\text{Effect 1}} (\obsY{1},A \con \widehat{\psi}_1) \big\} $ where $\Psi_{\text{Effect 1}}(\obsY{1},A \con \psi_1) = A( \obsY{1} - \psi_1 )$. 

To construct an estimator for $\psi_0^\true$, we leverage the moment equation \eqref{eq-ATT0-DR}. 
From \eqref{eq-propensitylikelihood}, we find the following odds is parametrized by $\eta_1$ and $\alpha_0$:
\begin{align*}
\frac{\pi_{1}^{\dagger}\left(  \obsY{1}  ,X\right)  }
{ 1-\pi_{1}^{\dagger} (  \obsY{1} ,X )  }
=
\exp 
\left\{ \left(1,X\T  \right) \eta_1 + \alpha_0\T  S_0 (\obsY{1},X) \right\}
\end{align*}
Additionally, under the parametric model in \eqref{eq-outcomelikelihood-baseline}, the baseline conditional density of $\obsY{1}$ given $A=0,X$ is proportional to $h_1(y , x \con \tau_{1},\gamma_{1})$. Therefore, the following function is parametrized by $(\tau_{1},\gamma_{1},\alpha_{0})$, denoted by $\xi(X \con \tau_{1},\gamma_{1},\alpha_0)$:
\begin{align*}
\xi(X \con \tau_{1},\gamma_{1},\alpha_{0})
=
\frac{E^{\dagger}\left[  \obsY{1}  \exp\left\{  \alpha
_{0}\T S_{0}( \obsY{1} ,X  )\right\}  |A=0,X \con \tau_{1}, \gamma_{1} \right]  }
{E^{\dagger} \left[  \exp\left\{  \alpha_{0}\T S_{0}(\obsY{1} ,X )\right\}
|A=0, X  \con \tau_{1}, \gamma_{1} \right] }
\end{align*}
Often times, $\xi(X \con \tau_{1},\gamma_{1},\alpha_{0})$ is represented in a simple form. 
For instance, if the baseline outcome is modeled as $\obsY{1} \cond (A=0,X) \sim N(\mu_1(A,X \con \tau_1, \gamma_1) ,\sigma_1^2(X \con \tau_1, \gamma_1))$ and the sufficient statistic is chosen as $S_0(\obsY{0},X)=(1,X\T ) \obsY{0}$, we obtain $\xi(X \con \tau_{1},\gamma_{1}, \alpha_{0}) = \mu_1(A,X, \con \tau_1, \gamma_1) + \sigma_1^2(X \con \tau_1,\gamma_1) \alpha_{0}\T  (1,X\T ) \T $.

Substituting the functions above in \eqref{eq-outcomelikelihood-baseline}, we construct a estimating function for $\psi_0^\true$ as follows:
\begin{align*}
& 
\Psi_{\text{Effect 0}} (\obsY{1},A,X \con \tau_1,\gamma_1,\eta_1,\alpha_0,\psi_0)
\\
&
=
(1-A)
\big[ \exp 
\left\{ \left(1,X\T  \right) \eta_1 + \alpha_0\T  S_0 (\obsY{1},X) \right\} \big]
\big\{
\obsY{1} - \xi(X \con \tau_{1},\gamma_{1},\alpha_{0})
\big\} 
+
A \xi(X \con \tau_{1},\gamma_{1},\alpha_{0}) - A \psi_0
\end{align*}
Then, we obtain an estimator of $\psi_0^\true$ as the solution to the estimating equation
\begin{align*}
0 = \AVER \Big\{ \Psi_{\text{Effect 0}} (\obsY{1},A,X \con \widehat{\tau}_1,\widehat{\gamma}_1,\widehat{\eta}_1,\widehat{\alpha}_0,\widehat{\psi}_0) \Big\}
\end{align*}
where $(\widehat{\tau}_1,\widehat{\gamma}_1,\widehat{\eta}_1,\widehat{\alpha}_0)$ are obtained from the previous estimating equations. 

Lastly, the average causal effect of treatment on the treated is obtained as $\widehat{\psi}=\Delta(\widehat{\psi}_0,\widehat{\psi}_1)$.

\subsubsection*{Statistical Properties of the Estimator}

For notational simplicity, let $\theta = (\tau_0,\gamma_0,\eta_0,\tau_1,\gamma_1,\eta_1,\alpha_0,\alpha_\OR,\alpha_\PS)$ and $O=(\obsY{0},\obsY{1},A,X)$. It is convenient to stack all estimating equations above, denoted by $\Psi$:
\begin{align}                   \label{eq-AllMoment}
\Psi (O \con \psi_0,\psi_1,\theta)
=
\begin{pmatrix}
\Psi_{\text{Effect 0}} (\obsY{1},A,X \con \tau_1,\gamma_1,\eta_1,\alpha_0,\psi_0)
\\
\Psi_{\text{Effect 1}} (\obsY{1},A,X \con \psi_1)
\\
\Psi_{t=0,\OR} (\obsY{0},A,X \con {\tau}_{0}, {\gamma}_{0}, {\alpha}_\OR) 
\\
\Psi_{t=0,\PS}(\obsY{0},A,X \con {\eta}_0, {\alpha}_\PS )
\\
\Psi_{\text{Odds Ratio}} (\obsY{0},A,X \con {\eta}_0, {\tau}_0, {\gamma}_0, {\alpha}_0)
\\
\Psi_{t=1,\OR} (\obsY{1},A,X \con {\tau}_{1}, {\gamma}_{1})
\\
\Psi_{t=1,\PS} (\obsY{1},A,X \con {\eta}_{1}, {\alpha}_{0})
\end{pmatrix}
\end{align}
The estimators of the parameters $(\psi_0,\psi_1,\theta)$ are obtained as the solution to the stacked estimating equation, i.e., $ 0
=
\AVER
\left\{
\Psi (O \con \widehat{\psi}_0 , \widehat{\psi}_1 , \widehat{\theta} )
\right\}$.

We can characterize the statistical properties of the estimator using the M-estimation theory (e.g., Chapter 5 of \citet{Vaart1996}, \citet{Stenfanski2002}). Under regularity conditions, we have the following asymptotic normality of the estimators:
\begin{align*}
& 
\sqrt{N}
\left\{
\big(
\widehat{\psi}_0 , \widehat{\psi}_1 , \widehat{\theta}
\big) \T 
-
\big(
\underline{\psi}_0 , \underline{\psi}_1 , \underline{\theta}
\big) \T 
\right\}
\stackrel{D}{\rightarrow}
N
\Big(
0
,
V_1^{-1} V_2 (V_1^{-1})\T  
\Big) \\
&
V_1
=
E \left. \left\{
\frac{ \partial \Psi(O \con \psi_0,\psi_1,\theta) }{\partial (\psi_0,\psi_1,\theta)\T  }
\right\} \right|_{ (\psi_0,\psi_1,\theta)=(\underline{\psi}_0,\underline{\psi}_1,\underline{\theta}) } 
\ , \ 
V_2 = 
E \left[
\{ \Psi(O \con \underline{\psi}_0,\underline{\psi}_1,\underline{\theta} ) \}
\{
\Psi(O \con \underline{\psi}_0,\underline{\psi}_1,\underline{\theta} ) \} \T 
\right]
\end{align*}
where $(\underline{\psi}_0 , \underline{\psi}_1 , \underline{\theta})$ is the probability limit of the estimators. From the law of large numbers, we have $\widehat{\psi}_1 = \AVER(A\obsY{1})/\AVER(A) \stackrel{P}{\rightarrow} \psi_1^\true$, implying $\underline{\psi}_1= \psi_1^\true$. Additionally, as shown in Section \ref{sec:supp:Ident}, we achieve $\underline{\psi}_0=\psi_0^\true$ if (i) $\underline{\alpha}_0 =\alpha_0$ and (ii) either $\underline{\eta}_1=\eta_1$ or $(\underline{\tau}_1, \underline{\gamma}_1) = (\tau_1,\gamma_1)$. Consequently, using the delta method, we obtain  $ \sqrt{N} \big(
\widehat{\psi} - \psi^\star
\big)
\stackrel{D}{\rightarrow}
N
\big(
0
,
\sigma^2
\big)$ where
\begin{align*}
&
\sigma^2 =
v(\psi_0^\true,\psi_1^\true)\T  
\{ V_1^{-1} V_2 (V_1^{-1})\T   \} 
v(\psi_0^\true,\psi_1^\true)  \ , \\
&
v (\psi_0,\psi_1)
=
\left(
\frac{\partial \Delta(\psi_0,\psi_1)}{ \partial \psi_0}
\ , \
\frac{\partial \Delta(\psi_0,\psi_1)}{ \partial \psi_1}
\ , \ 0  \ , \ \ldots \ , \ 0
\right)
\end{align*}
A consistent variance estimator of $\widehat{\psi}$ can be obtained by substituting the empirical analogs of $V_1$, $V_2$, $\psi_0^\true$, and $\psi_1^\true$, i.e., $\widehat{\sigma}^2 = v(\widehat{\psi}_0,\widehat{\psi}_1)\T  \{ \widehat{V}_1^{-1} \widehat{V}_2 (\widehat{V}_1^{-1})\T   \} v(\widehat{\psi}_0,\widehat{\psi}_1)$ where
\begin{align*}
&
\widehat{V}_1 
=
\AVER \bigg\{
\frac{ \partial \Psi(O \con \psi_0,\psi_1,\theta ) }{\partial (\psi_0,\psi_1,\theta)\T  } 
\bigg\}\bigg|_{ (\psi_0,\psi_1,\theta) = ( \widehat{\psi}_0,\widehat{\psi}_1,\widehat{\theta})}
\ , \ 
\widehat{V}_2
=
\AVER \left[
\{ \Psi(O \con \widehat{\psi}_0,\widehat{\psi}_1,\widehat{\theta}) \}
\{ \Psi(O \con \widehat{\psi}_0,\widehat{\psi}_1,\widehat{\theta}) \}\T 
\right]
\end{align*}
Confidence intervals can be constructed leveraging the asymptotic normality of $\widehat{\psi}$ and the consistent variance estimator $\widehat{\sigma}^2$. A $100(1-\alpha)$\% confidence interval for $\psi^\true$ is given as 
\begin{align*}
\left( \widehat{\psi} - z_{1-\frac{\alpha}{2}} \frac{ \widehat{\sigma}}{\sqrt{N}}, 
\widehat{\psi} + z_{1-\frac{\alpha}{2}} \frac{ \widehat{\sigma}}{\sqrt{N}} \right)
\end{align*}
where $z_{\alpha}$ is the $100\alpha$th percentile of the standard normal distribution.

\subsection{Details on Estimation with Discretized Odds Ratio Functions}      \label{sec:supp:Diagnosis}

In this Section, we explore a straightforward approach to conduct model diagnostics and to potentially improve the robustness of estimation of the log-odds ratio function which plays a key role in the proposed approach. In the previous Section this function was specified to follow the parametric form $\beta_0(y)=  \alpha_0 y$, a linear function of $y$ which may be overly restrictive when $Y$ is continuous. A straightforward approach to improve robustness entails modeling the odds ratio function as a discrete function of the outcome, analogous to a histogram estimator of a density.  
We proceed by binning the outcome into $M$ bins based on the $100m/M$th percentiles of the pre-exposure outcome where $m=1,2,\ldots,M-1$. In the application, we take $M=10$ bins and define the discretized outcome $B(\obsY{t}) \in \{1,2,\ldots,M\}$ where $B(\obsY{t})=m$ if $\obsY{t}$ belongs to the $m$th bin. Thus, for $M=10$, the outcome is discretized based on the deciles of the pre-exposure outcome, and the pre-exposure discretized outcome $B(\obsY{0})$ is thus approximately distributed uniformly over the support $\{1,\ldots,M\}$ by design. Next, we specify the odds ratio function as $ \beta_0(y) = \exp \{ \alpha_{0,B(y)} \} $ where $\alpha_{0,m}$ $(m=1,\ldots,M)$ is a vector of parameters of the odds ratio function associated with the outcome over the $m$th bin. Without loss of generality, we set the first bin as reference bin, and consequently $\alpha_{0,1}$ is a vector of zeros. We refer to this odds ratio function as the discretized odds ratio function. Together with the discretized odds ratio, we specify a polytomous logistic regression to model the discretized outcome densities at both $t=0,1$.

Now, we discuss details on estimation of the discretized odds ratio function. We denote the discretized odds ratio function  as $\beta_0(y,x)
=
\exp 
\big\{ 
z\T  \alpha_{0,B(y)}
\big\}$ where $Z$ is a function of $X$. In the Zika virus application, we use $Z=1$ with $p=dim(Z)=1$.

We first redefine $\Psi_{t=0,\OR}$ in \eqref{eq-AllMoment}. We estimate the parameters $\alpha_{0,m}$ $(m=2,\ldots,M)$ from multinomial logistic regressions which are equivalent to solving the following system of the estimating equations: $\AVER \big\{ \Psi_{t=0,\OR} (\obsY{0},A,X \con \tau_0, \alpha_{\text{OR}}) \big\} = 0$ where $\tau_0 = (\tau_{0,2},\ldots,\tau_{0,M})$ and $\alpha_{\text{OR}}=(\alpha_{\text{OR},2},\ldots,\alpha_{\text{OR},M})$ with
\begin{align}
&
\Psi_{t=0,\OR}(\obsY{0},A,X \con \tau_0, \alpha_{\text{OR}})
=
\begin{pmatrix}
\Psi_{t=0,\OR,2}(\obsY{0},A,X \con \tau_{0}, \alpha_{\text{OR}} ) 
\\
\vdots 
\\
\Psi_{t=0,\OR,M}(\obsY{0},A,X \con \tau_{0}, \alpha_{\text{OR}} ) 
\end{pmatrix}    
\label{eq-MultNomi}
\\
\nonumber
&
\Psi_{t=0,\OR,k}(\obsY{0},A,X \con \tau_{0}, \alpha_{\text{OR}} ) 
\\
\nonumber
&
=
\begin{pmatrix}
1 \\ X \\ A \cdot Z
\end{pmatrix}
\left[
\ind \big\{ B(\obsY{0})=k \big\}
- 
\Pr \big\{ B(\obsY{0})=k \cond A,X \con \tau_{0}, \alpha_{\text{OR}} \big\} 
\right]
\ , \
k=2,\ldots,M
\end{align}
Note that $\Psi_{t=0,\OR,k}$ is $(1+\dim(X)+p)$-dimensional, and thus, equation \eqref{eq-MultNomi} is $(M-1)(1+\dim(X)+p)$-dimensional. Additionally, we can represent the conditional probability as 
\begin{align}     \label{eq-MultNomi2}
\Pr \big\{ B(\obsY{0})=k \cond A,X \con \tau_{0}, \alpha_{\text{OR}} \big\} 
=
\frac{ \exp \big\{ (1,X\T ) \tau_{0,k} +
A \cdot Z\T  \alpha_{\text{OR},k} \big\}
}{ 1 + \exp \big[ \sum_{m=2}^{M}
\big\{
(1,X\T ) \tau_{0,m} +
A \cdot Z\T 
\alpha_{\text{OR},m} \big\}
\big] }
\end{align}

We specify the baseline outcome likelihood at time 1 using the same $\Psi_{t=1,\OR}$ in \eqref{eq-AllMoment}, which is used to estimate $(\tau_1,\gamma_1)$. Next, to evaluate $\Psi_{\text{Effect 0}}$ in \eqref{eq-AllMoment}, we need to evaluate $\xi(X)$ which is given by
\begin{align*}
    \xi(X)
    =  \frac{E\left[
\obsY{1}  \exp\left\{  \beta_{0}\left(  \obsY{1} , X  \right)
\right\}  |A=0,X\right]  }{E\left[  \exp\left\{  \beta_{0}\left(  \obsY{1} , X
\right)  \right\}  |A=0,X\right]  }
\end{align*}
 of which close-form representation can be obtained.

For instance, if $\potY{0}{1} \cond (A=0,X) \sim N(\mu_1(0,X \con \tau_{1}), \sigma_1^2(X \con \gamma_{1}) )$, then $\xi(X)$ is represented as
\begin{align*}
&
\xi(X \con \alpha_0, \tau_1, \gamma_1)
=
\frac{ 
\displaystyle{
\sum_{m=1}^{M}
\Big[
\exp\{ Z\T  \alpha_{0,m} \}
\big\{ 
\mu_1(0,X \con \tau_1)     
Q_{1,m}(X \con \tau_1, \gamma_1)
-
\sigma_1^2(X \con \gamma_1)
Q_{2,m}(X \con \tau_1, \gamma_1)
\big\}
\Big]
}
}
{
\displaystyle{
\sum_{m=1}^{M}
\Big[
\exp\{ Z\T  \alpha_{0,m} \}
Q_{1,m}(X \con \tau_1, \gamma_1)
\Big]
}
}
\\
&
Q_{1,m}(X \con \tau_1, \gamma_1)
=
\Phi( u_m \con \mu_1(0,X \con \tau_1), \sigma_1^2(X \con \gamma_1) ) - \Phi( \ell_m \con \mu_1(0,X \con \tau_1), \sigma_1^2(X \con \gamma_1) )
\ , \\
&
Q_{2,m}(X \con \tau_1, \gamma_1)
=
\phi( u_m \con \mu_1(0,X \con \tau_1), \sigma_1^2(X \con \gamma_1) ) - \phi( \ell_m \con \mu_1(0,X \con \tau_1), \sigma_1^2(X \con \gamma_1) )
\end{align*}
Here $\Phi(\cdot \con \mu, \sigma^2)$ and $\phi(\cdot \con \mu, \sigma^2)$ are the cumulative distribution function and the density function of $N(\mu,\sigma^2)$, respectively. As mentioned before, we use $\xi(X \con \alpha_0,\tau_1, \gamma_1)$ to evaluate $\Psi_{\text{Effect 0}}$ in \eqref{eq-AllMoment}.

Next, we update the estimating equations for the propensity score. We find 
\begin{align*}
\log\frac{\pi_{t}\left(  y,x\right)  }{1-\pi_{t}\left(  y,x\right)  }=\left(
1,x\T \right)  \eta_{t} + \log \beta_0(y,x)
\end{align*}
where $\beta_0(y,x)= \exp \big\{ z\T  \alpha_{0,\PS,B(y)} \big\}$. Therefore, $\Psi_{t=0,\text{PS}}$ in equation \eqref{eq-AllMoment} is updated as
\begin{align*}
\Psi_{t=0,\text{PS}}(\obsY{0},A,X \con \eta_0,\alpha_\PS )
=
\begin{pmatrix}
1 \\ X \\ \ind \big\{ B(\obsY{0}) = 2 \big\} Z \\ \vdots  \\ \ind \big\{ B(\obsY{0}) = M \big\} Z
\end{pmatrix}
\left[
A - \text{expit} \Big[ (1,X\T ) \eta_0 + 
\sum_{m=2}^{M} \Big[  \ind \big\{ B(\obsY{0} ) = m \big\} Z\T  \alpha_{0,\PS,m} \Big]
\Big]
\right]
\end{align*}
$\Psi_{t=1,\text{PS}}$ in \eqref{eq-AllMoment} has the same form except that the odds ratio has the discrete form:
\begin{align*}
\Psi_{t=1,\PS} (\obsY{1},A,X \con \eta_{1},\alpha_{0} )
= \begin{pmatrix}
1 \\ X
\end{pmatrix}
\big[  \left(  1-A\right)  \left[  1+\exp\left\{  \left( 1,X\T \right)  \eta_{1}+ Z\T  \alpha_{0,B(\obsY{1})} \right\}  \right] -1 \big]
\end{align*}

For the doubly robust estimator, we use the following $(M-1) p$-dimensional moment equation as $\Psi_{\text{Odds Ratio}}$ in \eqref{eq-EE_OddsRatio} instead of \eqref{eq-OddsRatioMomentEquation} with $\alpha_0=(\alpha_{0,2},\ldots,\alpha_{0,M})$:
\begin{align*}
&
\Psi_{\text{Odds Ratio}} (\obsY{0},A,X \con \eta_0, \tau_0, \alpha_{0})
\\
&
: =
\left[  A
-
\frac{ \exp\{ (1,X\T ) \eta_0 \} }{1+\exp\{ (1,X\T ) \eta_0 \}}
\right]
\exp \big\{ - A Z\T  \alpha_{0,B(\obsY{0})} \big\}
\begin{bmatrix}
Z\T  \big[
\ind \{ B(\obsY{0})= 2 \}-
\Pr \big\{ B(\obsY{0})= 2  \cond A=0, X \con \tau_0 \big\}
\big]
\\
\vdots 
\\
Z\T  \big[
\ind \{ B(\obsY{0})= M \}-
\Pr \big\{ B(\obsY{0})= M  \cond A=0, X \con \tau_0  \big\}
\big]
\end{bmatrix}
\end{align*}
Here $\Pr \big\{ B(\obsY{0})= k  \cond A=0, X \con \tau_0 \big\}$ is evaluated from \eqref{eq-MultNomi2}. With these updated estimating equations, we can obtain the estimator of the causal effect from \eqref{eq-MultNomi2}.

\subsection{Application: Model Diagnostics by Using Discretized Odds Ratio Functions}      \label{sec:supp:Analysis Discrete}

We provide additional data analysis results by performing model diagnostics using the discretized odds ratio function presented in the previous Section \ref{sec:supp:Diagnosis}. We specify the odds ratio function as $ \beta_0(y,x) = \exp \{ (1,x\T ) \alpha_{0,B(y)} \} $ where $\alpha_{0,m}$ $(m=1,\ldots,M)$ is a vector of parameters of the odds ratio function associated with the outcome over the $m$th bin. Without loss of generality, we set the first bin as reference bin, and consequently $\alpha_{0,1}$ is a vector of zeros. We refer to this odds ratio function as the discretized odds ratio function. Together with the discretized odds ratio, we specify a polytomous logistic regression to model the discretized outcome densities at both $t=0,1$.

Using the discretized odds ratio function for the data analysis yields results summarized in Table \ref{Tab-2-Additional} together with the results in the main paper based on the log-linear odds ratio model, i.e., $\beta_0(y,x) = \alpha_0 \T S_0(y,x)$ with $S_0(y,x)=(y,yx\T)$; these specifications are referred to as UDiD (Discretized) and UDiD (Log-linear), respectively. We find that the estimates from the discretized odds ratio function are generally consistent with those from the log-linear odds ratio function. However, the estimates from the discretized odds ratio function are much closer to each other, demonstrating less model dependence as well as robustness, with substantially tighter confidence intervals. In addition, the estimates under PT are of a similar value as the estimates under OREC obtained from the discretized odds ratio function. Nonetheless, all effect estimates are significant at 0.05 level and, compared to the crude estimate of 3.384, the negative effect estimates suggest the presence of substantial confounding bias. The additional data analysis results corroborate the finding in the main paper that the Zika virus outbreak led to a decline in the number of births in Brazil, and causal conclusions are not particularly sensitive to the choice of identifying assumptions and/or specification of the odds ratio function.

\begin{table}[!htp]
\renewcommand{\arraystretch}{1.2} \centering
\setlength{\tabcolsep}{4pt}
\begin{tabular}{|cc|ccc|}
\hline
\multicolumn{2}{|c|}{\multirow{2}{*}{Estimator}}                                                                                            & \multicolumn{3}{c|}{Statistic}                                                \\ \cline{3-5} 
\multicolumn{2}{|c|}{}                                                                                                                      & \multicolumn{1}{c|}{Estimate} & \multicolumn{1}{c|}{SE}    & $95\%$ CI        \\ \hline
\multicolumn{1}{|c|}{\multirow{3}{*}{\begin{tabular}[c]{@{}c@{}}UDiD under OREC\\ (Log-linear)\end{tabular}}}  & GLM-based                  & \multicolumn{1}{c|}{-1.487}   & \multicolumn{1}{c|}{0.340} & (-2.153, -0.821) \\ \cline{2-5} 
\multicolumn{1}{|c|}{}                                                                                         & Propensity score weighting & \multicolumn{1}{c|}{-0.831}   & \multicolumn{1}{c|}{0.377} & (-1.570, -0.091) \\ \cline{2-5} 
\multicolumn{1}{|c|}{}                                                                                         & Doubly-robust              & \multicolumn{1}{c|}{-0.974}   & \multicolumn{1}{c|}{0.342} & (-1.645, -0.303) \\ \hline
\multicolumn{1}{|c|}{\multirow{3}{*}{\begin{tabular}[c]{@{}c@{}}UDiD under OREC\\ (Discretized)\end{tabular}}} & GLM-based                  & \multicolumn{1}{c|}{-1.012}   & \multicolumn{1}{c|}{0.259} & (-1.519, -0.505) \\ \cline{2-5} 
\multicolumn{1}{|c|}{}                                                                                         & Propensity score weighting & \multicolumn{1}{c|}{-1.080}   & \multicolumn{1}{c|}{0.244} & (-1.559, -0.601) \\ \cline{2-5} 
\multicolumn{1}{|c|}{}                                                                                         & Doubly-robust              & \multicolumn{1}{c|}{-0.964}   & \multicolumn{1}{c|}{0.262} & (-1.476, -0.451) \\ \hline
\multicolumn{1}{|c|}{\multirow{3}{*}{\begin{tabular}[c]{@{}c@{}}Standard DiD\\ under PT\end{tabular}}}         & GLM-based                  & \multicolumn{1}{c|}{-1.171}   & \multicolumn{1}{c|}{0.151} & (-1.467, -0.875) \\ \cline{2-5} 
\multicolumn{1}{|c|}{}                                                                                         & Propensity score weighting & \multicolumn{1}{c|}{-1.091}   & \multicolumn{1}{c|}{0.143} & (-1.371, -0.811) \\ \cline{2-5} 
\multicolumn{1}{|c|}{}                                                                                         & Doubly-robust              & \multicolumn{1}{c|}{-1.124}   & \multicolumn{1}{c|}{0.159} & (-1.435, -0.813) \\ \hline
\end{tabular}
\caption{\footnotesize Summary of Data Analysis. Values in ``Estimate'' column represent the average causal effect of the Zika outbreak on the birth rate within the PE region, i.e., the difference between PE's observed average birth rate to a forecast of what it would have been had the Zika outbreak been prevented. Values in ``SE'' and ``95\% CI'' columns represent the standard errors associated with the estimates and the corresponding 95\% confidence intervals, respectively. The reported values are expressed as births per 1,000 persons.}
\label{Tab-2-Additional}

\end{table}

\subsection{Sensitivity Analysis}       \label{sec:supp:sensitivity}

We provide details on a sensitivity analysis to the OREC assumption. 
Consider that the odds ratio function $\beta_{1}\left(  y,x\right)$ is set to equal: \[
\beta_{1}\left(  y,x\right)=\beta_{0}\left(  y,x\right)+\Delta(y,x);
\]
where $\Delta(y,x)$ is a user-specified non-identifiable sensitivity function that encodes a potential departure from OREC.  
For instance, one might specify $\Delta(y,x)= d \times T(y,x)$ for a function $T$ and a scalar $d$ varying over a grid of values in the interval $(-\omega;\omega)$ with zero corresponding to OREC; here $\omega$ and $T$ are user-specified, and a simpler form of $T$ is often preferred in practice, say $T(y,x) = y$. 
For each value of $d$ in the interval, one would then proceed by repeating the proposed analyses and reporting various updated estimates of the causal effect of interest thus providing an evaluation of the sensitivity of inferences to possible violations of the OREC condition. 

We conduct sensitivity analysis to the OREC assumption for the data application in Section \ref{sec:Data}. Let the pre-exposure odds ratio is given as $\exp\big\{ \alpha_0\T  S(y,x) \big\}$. Then, as discussed above, the post-exposure odds ratio is assumed to be represented as $\exp \big\{ \alpha_0\T  S(y,x) + d y \big\}$ where $d$ is a scalar. For interpretability, $d$ can be reparametrized as $d = d'/\sigma_{Y}$ where $d'$ is a scalar and $\sigma_{Y}$ is a reasonable candidate for the (conditional) variance of the post-exposure outcome; in this case, $d \obsY{1}$ roughly has a standard deviation of $d'$. For a given sensitivity parameter $d$ (or $d'$), the average causal effect of treatment on the treated of the form $\Delta(\psi_0,\psi_1)$ can be obtained from the following estimating equation; see the previous Section \ref{sec:supp:Estimation} for details:
\begin{align*}
\Psi_D (O \con \psi_0,\psi_1,\theta)
=
\begin{pmatrix}
\Psi_{\text{Effect 0}} (\obsY{1},A,X \con \tau_1,\gamma_1,\eta_1,\alpha_0 + d \vec{e}_Y,\psi_0)
\\
\Psi_{\text{Effect 1}} (\obsY{1},A,X \con \psi_1)
\\
\Psi_{t=0,\OR} (\obsY{0},A,X \con {\tau}_{0}, {\gamma}_{0}, {\alpha}_\OR) 
\\
\Psi_{t=0,\PS}(\obsY{0},A,X \con {\eta}_0, {\alpha}_\PS )
\\
\Psi_{\text{Odds Ratio}} (\obsY{0},A,X \con {\eta}_0, {\tau}_0, {\gamma}_0, {\alpha}_0)
\\
\Psi_{t=1,\OR} (\obsY{1},A,X \con {\tau}_{1}, {\gamma}_{1})
\\
\Psi_{t=1,\PS} (\obsY{1},A,X \con {\eta}_{1}, {\alpha}_{0} + d \vec{e}_Y)
\end{pmatrix}
\end{align*}
where $\vec{e}_Y$ is a vector satisfying $\vec{e}_Y\T S_0(y,x)=y$; if $S_0(y,x) = (1,x\T )y$, then $\vec{e}_Y=(1,0,\ldots,0)\T $. We then vary $d$ (or $d'$) from a user-specified interval, say $d \sigma_{Y}=d' \in [-2,2]$ and check how the estimated effects and corresponding confidence intervals change. The original analysis is said to be robust to the OREC assumption if the causal conclusions obtained from the original analysis do not change even at large sensitivity parameters. In contrast, if the established causal conclusions change at a small sensitivity parameter (e.g., effect becomes non-significant at a given level and/or effect becomes zero), the original analysis can be said to be sensitive to the OREC assumption.

\begin{figure}[!htb]
\centering
\includegraphics[width=1\textwidth]{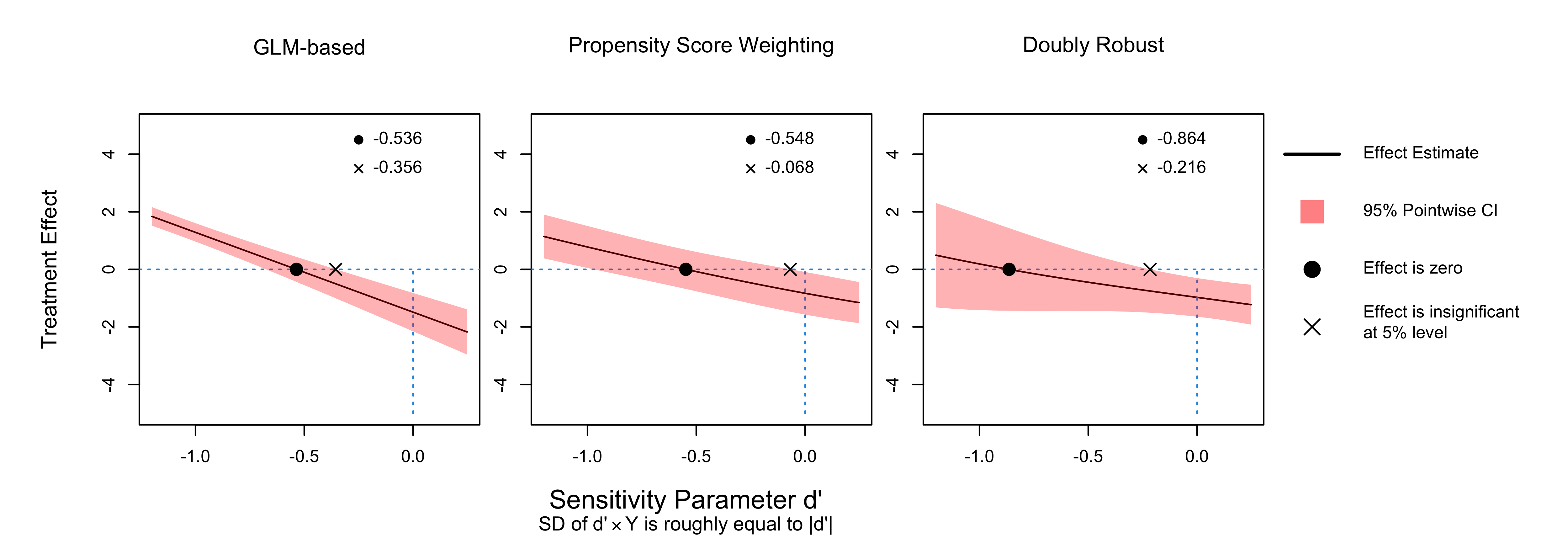}
\caption{\footnotesize Sensitivity Analysis. Plots in the left, middle, and right columns correspond to the GLM-based, propensity score weighting, and doubly robust UDiD estimators, respectively. The blue dotted horizontal and vertical lines visually guide the zero effect and zero sensitivity parameter, respectively.}
\label{fig-2-Additional}
\end{figure}

We apply the sensitivity analysis to the real-world application in Section \ref{sec:Data}. Figure \ref{fig-2-Additional} visually summarizes the sensitivity analysis results of the application. 
We find that the propensity score weighting UDiD estimate is the most sensitive to the violation of OREC. The doubly robust UDiD estimate becomes non-significant at $d'=-0.216$, which is larger than the corresponding value of the GLM-based UDiD estimate with a value of $d'=-0.356$, but it becomes zero at $d'=-0.864$ which is smaller than values obtained from the other estimates. 

We next discuss how to interpret the sensitivity values. For simplicity, we only focus on the GLM-based UDiD approach where the estimate is no longer significant at $d'=-0.356$. Since the $d'\times \obsY{1}$ can be approximated to a standard normal distribution, the distribution of the ratio between the post-exposure odds ratio to the pre-exposure odds ratio can be approximated to $\exp\big( d' \cdot Z ) $ where $Z$ is a standard normal distribution. Based on the Monte Carlo simulation at $d'=-0.35$ where the GLM-based UDiD estimate is still significant at 5\% level, roughly 56\% of the post-exposure odds ratio is greater or less than the pre-exposure odds ratio by 20\%. This indicates that, even though more than half of the observations have notably different pre- and post-exposure odds ratio values (here, 20\% change is considered as a notable difference), the effect estimate remains significant and therefore the causal conclusion remains valid. Based on this investigation, we believe that the GLM-based UDiD estimate is fairly robust to the violation of OREC. The sensitivity parameters of the other estimates can be interpreted in a similar manner, but we omit the details here. 

\newpage

\subsection{Data Generating Processes for Odds Ratio Equi-confounding} \label{sec:supp:DGP}

We first establish \eqref{eq-model OREC} is compatible with OREC. Recall that \eqref{eq-model OREC} is
\begin{align*}
    &
    \potY{0}{t} \indep A \cond U_t \ , \ t=0,1 
    \quad , \quad 
    A \cond (U_1= u) \stackrel{D}{=} A \cond (U_0=u) \ , \ \forall u
    \ , \\
    &
    U_1 \cond (A=0,Y_1=y) \stackrel{D}{=} U_0 \cond (A=0,Y_0=y)
    \ , \ \forall y
     \ .
\end{align*}
For time $t=0,1$, the odds of treatment at $\potY{0}{t}=y$ is
\begin{align}
&
\frac{ 
\Pr(A=1 \cond \potY{0}{t} = y)
}{ 
\Pr(A=0 \cond \potY{0}{t} = y)
}	
\nonumber
\\
&
=
\int 
\frac{
P(\potY{0}{t} = y, A=1, U_t=u)
}{ 
P (\potY{0}{t}=y , A=0)
} 
\, du
\nonumber
\\
&
=
\int 
\frac{
P(\potY{0}{t} = y, A=1, U_t=u)
}{ 
P(\potY{0}{t} = y, A=0, U_t=u)
}
\frac{ 
P (\potY{0}{t}=y , A=0, U_t=u)
}{
P (\potY{0}{t}=y , A=0)
} 
\, du 
\nonumber
\\  
&
=
\int 
\frac{
\Pr(A=1 \cond \potY{0}{t} = y, U_t=u)
}{ 
\Pr(A=0 \cond \potY{0}{t} = y, U_t=u)
}
P (U_t = u \cond \potY{0}{t}=y, A=0 )
\, du
\nonumber
\\  
&
=
\int 
\frac{
\Pr(A=1 \cond  U_t=u)
}{ 
\Pr(A=0 \cond U_t=u)
}
P (U_t = u \cond \potY{0}{t}=y, A=0 )
\, du \ .
\label{eq-model OREC cond1}
\end{align}
The first four lines are trivial. 
The fifth line is from the first condition of \eqref{eq-model OREC}. From the third condition of \eqref{eq-model OREC}, we find that 
\begin{align} \label{eq-model OREC cond3}
    P (U_1 = u \cond \potY{0}{1}=y, A=0 )
    &
    = P (U_1 = u \cond Y_1=y, A=0 ) 
    \nonumber
    \\
    &
    = P (U_0 = u \cond Y_0=y, A=0 ) = P (U_0 = u \cond \potY{0}{0}=y, A=0 )
     \ .
\end{align}
Therefore, we establish that
\begin{align*}
\frac{ 
\Pr(A=1 \cond \potY{0}{0} = y)
}{ 
\Pr(A=0 \cond \potY{0}{0} = y)
}	
&
=
\int 
\frac{
\Pr(A=1 \cond \potY{0}{t} = y, U_0=u)
}{ 
\Pr(A=0 \cond \potY{0}{t} = y, U_0=u)
}
P (U_0 = u \cond \potY{0}{t}=y, A=0 )
\, du
\\
&
=
\int 
\frac{
\Pr(A=1 \cond \potY{0}{1} = y, U_1=u)
}{ 
\Pr(A=0 \cond \potY{0}{1} = y, U_1=u)
}
P (U_1 = u \cond \potY{0}{t}=y, A=0 )
\, du
\\
&
=
\frac{ 
\Pr(A=1 \cond \potY{0}{1} = y)
}{ 
\Pr(A=0 \cond \potY{0}{1} = y)
}	\ .
\end{align*}
The first and third identities are from \eqref{eq-model OREC cond1}. The second identity is from \eqref{eq-model OREC cond3} and the second condition of \eqref{eq-model OREC}. Therefore, this implies that the odds ratio is the same over time:
\begin{align*} 
\beta_{0} (y) 
& =
\log \bigg\{
\frac{ 
\Pr(A=1 \cond \potY{0}{0} = y)
}{ 
\Pr(A=0 \cond \potY{0}{0} = y)
}	
\frac{ 
\Pr(A=0 \cond \potY{0}{0} = 0)
}{ 
\Pr(A=1 \cond \potY{0}{0} = 0)
}	
\bigg\}
\\
&
=
\log \bigg\{
\frac{ 
\Pr(A=1 \cond \potY{0}{1} = y)
}{ 
\Pr(A=0 \cond \potY{0}{1} = y)
}	
\frac{ 
\Pr(A=0 \cond \potY{0}{1} = 0)
}{ 
\Pr(A=1 \cond \potY{0}{1} = 0)
} 
\bigg\}
=\beta_{1}\left(  y,x\right)
  \ .
\end{align*}

As a specific example of a continuous outcome, we can consider the following data generating process for $t=0,1$:
\begin{align} \label{eq-CiCType-Conti}
\potY{0}{t} = U_t + \epsilon_t
\ , \
\left\{
\begin{array}{l}
U_t \cond (A=0) \sim N(\mu_U,\sigma_U^2)
\\
U_t \cond (A=1) \sim p_U(u \cond A=1)
\end{array}
\right.
\ , \ 
\epsilon_t \cond (A,U_t) \sim N(0,\sigma_\epsilon^2)
\ .
\end{align}
The first condition of \eqref{eq-model OREC} is trivially satisfied because $\potY{0}{t} \cond (A,U_t) \sim N(U_t,\sigma_\epsilon^2)$ does not depend on $A$. In addition, from straightforward algebra, we obtain the following result for $t=0,1$:
\begin{align*}
&
\potY{0}{t} \cond (A=0,U_t) \sim N \big( U_t , \sigma_\epsilon^2 \big) 
\text{ with the density function } 
\phi (y-U_t \con \sigma_\epsilon^2)
\\
&
\potY{0}{t} \cond (A=1,U) \sim p_Y(y \cond A=1,U_t) 
\propto p_U(U_t \cond A=1) 
\phi (y-U_t \con \sigma_\epsilon^2) \ .
\end{align*}
where $\phi(z \con \sigma^2)$ is the density function of $N(0,\sigma^2)$. 

The second condition of \eqref{eq-model OREC} is satisfied because $P(A \cond U_t=u)$ does not depend on time as follows:
\begin{align*}
    & P(U_t=u,A=a)
    = \Pr (A=a) P(U_t =u \cond A=a)
    \\
    & \Rightarrow 
    P(U_t=u)
    =
    \Pr (A=1) p_U(u \cond A=1)
    +
    \Pr (A=0) \phi(u - \mu_U \con \sigma_U^2)
 \\
    & \Rightarrow 
    P(A \cond U_t=u)
    =
    \frac{
    A \cdot \Pr (A=1) p_U(u \cond A=1)
    +
    (1-A) \cdot \Pr (A=0) \phi(u - \mu_U \con \sigma_U^2)
    }
    { \Pr (A=1) p_U(u \cond A=1)
    +
    \Pr (A=0) \phi(u - \mu_U \con \sigma_U^2) } 
\end{align*}

The third condition of \eqref{eq-model OREC} is satisfied with
\begin{align*}
    U_t \cond (A=0,\potY{0}{t}=y) \sim N \left( \frac{ \mu_U/\sigma_U^2 + y/\sigma_\epsilon^2 }{1/\sigma_U^2+1/\sigma_\epsilon^2} 
, \frac{1}{1/\sigma_U^2+1/\sigma_\epsilon^2} \right) \ .
\end{align*}

For a data generating process for a binary outcome that is compatible with OREC, we may consider the following model, which is first introduced in the Supplementary Material of \citet{DiD_OREC2022}:
\begin{align*}
&
\potY{0}{0} = \ind \big(  b_0  + U _0 \geq 0 \big)
\ , 
&&
\potY{0}{1} = \ind \big(  b_1 + U_1 \geq 0 \big) \ ,
\end{align*}
Here, $(U_0,U_1)$ are latent variables following $U_t \cond A \sim \text{Logistic}( \nu(A) ,\sigma_U )$, and $(U_0,U_1)$ are allowed to have an arbitrary correlation structure. The scalars $b_t$ parametrize the unit's time-specific base level. The treatment-free potential outcomes are discretized values in the indicator functions. Then, after some algebra, we find
\begin{align*}
&
\potY{0}{0} \cond A
\sim \text{Ber} \Bigg(  \text{expit} \bigg( 
\frac{  b_0  + \nu(A) }{\sigma_U}  \bigg)
\Bigg)
\ , \
&
\potY{0}{1} \cond A
\sim
\text{Ber} \Bigg(
\text{expit} \bigg( 
\frac{  b_1 + \nu(A) }{\sigma_U}  \bigg) 
\Bigg)
\ .
\end{align*}
which satisfies OREC with $\beta_0(y) = \sigma_U^{-1} \big\{ \nu(1) - \nu(0) \big\} y$. 

We refer the readers to \citet{DiD_OREC2022} for additional data generating processes for continuous and count outcomes.

\subsection{Assessment of Covariate Distribution Invariance} \label{sec:supp:Cov Invariance}

The third condition of \eqref{eq-model OREC} states that the distribution of $U_t$ conditional on $A=0$ and $Y_t$ is stable over time. Therefore, to validate the plausibility of \eqref{eq-model OREC}, an analyst might inspect the extent to which the distribution of measured covariates $X$ remains stable in time conditional on $A=0$ and $Y_t$.  While these empirical checks are not formal tests, they are intuitive and easy to implement sanity checks as to the appropriateness of the condition (at least with respect to observed covariates) as the assumption may seem more reasonable if they are found to hold for observed covariates. We have now included versions of these simple empirical checks in the context of the Ziva virus application.

We denote the three covariates, the log population, population density, and proportion of females, of a municipality by $X_{1}$, $X_{2}$, and $X_{3}$, respectively. Likewise, we denote the birth rate at time $t=0,1$ of a municipality by $Y_{t}$. Recall that we only use untreated municipalities, i.e., municipalities in Rio Grande do Sul (RS), because it suffices to check the condition only among control units.

\begin{figure}[!htb]
\centering
\includegraphics[width=0.75\textwidth]{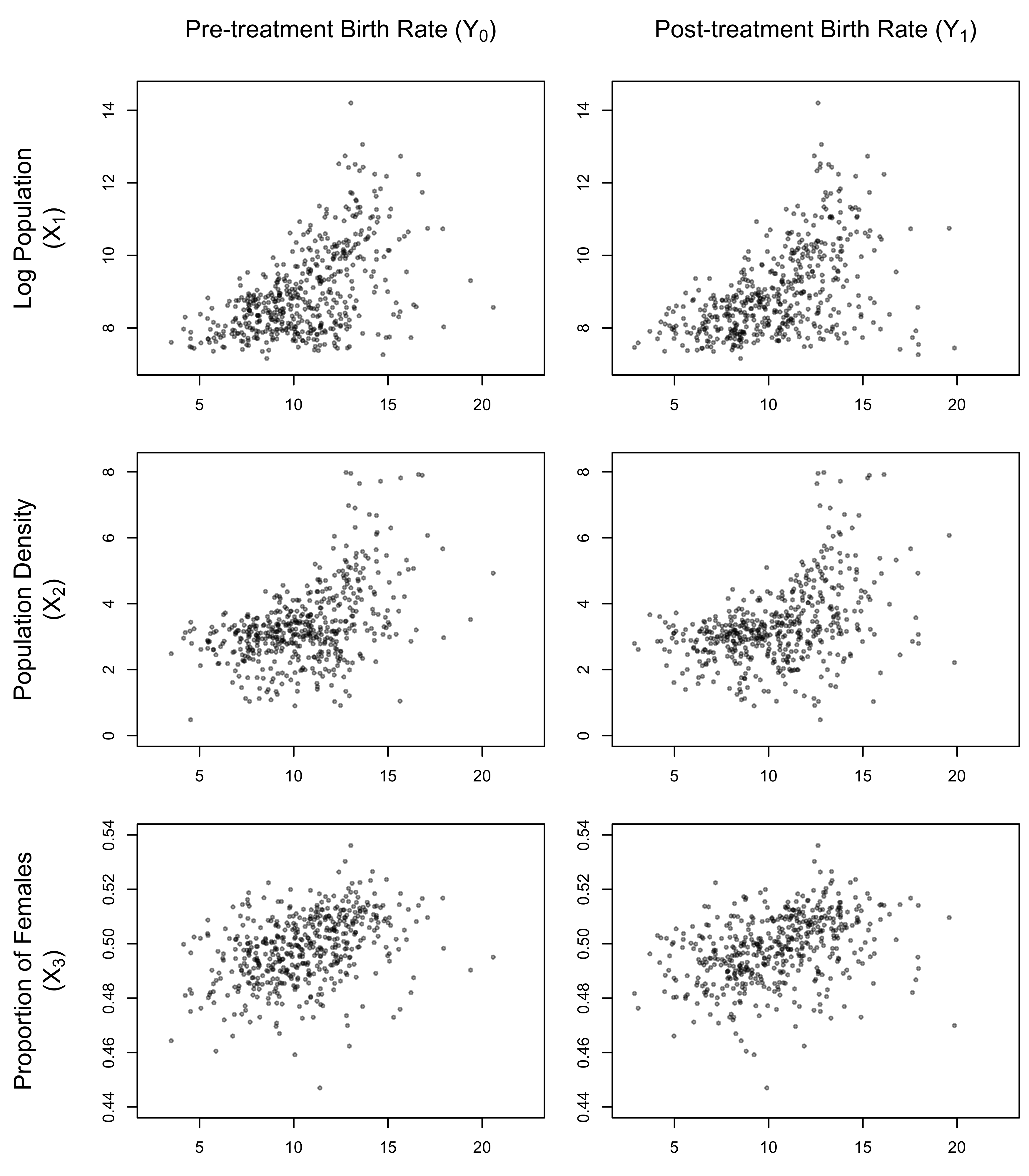}
\caption{\footnotesize A Graphical Summary of the Relationship between $Y_{t}$ and $X_j$. Each row corresponds to one of the three covariates (log population, population density, and proportion of females),  respectively. The left and right columns correspond to the pre- and post-exposure outcomes (birth rate), respectively. The $x$- and $y$-axes display the values of the outcomes and covariates, respectively.}
\label{fig-Cov Invariance}
\end{figure}

First, we visually assess whether the relationship between $X_{j}$ ($j=1,2,3$) and $Y_{t}$ does not dramatically change over time. Figure \ref{fig-Cov Invariance} graphically summarizes the empirical joint distribution of $Y_{t}$ and $X_{j}$. Upon visual examination, we observe that the associations between covariates and outcomes remain relatively stable across time periods.

Next, we perform statistical testing procedures based on a parametric model. Specifically, we consider a regression model where $X_{j}$ is considered as a response variable, and the intercept, $Y_t$, and $Y_t^2$ are considered as explanatory variables, i.e., 
\begin{align*}
    &
    X_j = \beta_{00j} + \beta_{01j} Y_0 + \beta_{02j} Y_0^2 + \epsilon_{0j}
    \ , 
    &&
    X_j = \beta_{10j} + \beta_{11j} Y_1 + \beta_{12j} Y_1^2 + \epsilon_{1j} \ .
\end{align*}
The quadratic terms are included to account for nonlinear relationships shown in Figure \ref{fig-Cov Invariance}. 
We then conduct statistical tests where null hypotheses are $H_{0,kj}: \beta_{0kj} = \beta_{1kj}$ for $k=1,2,3$. The null hypothesis is rejected at 5\% level if the corresponding Wald statistic is greater than $\chi_{1,0.95}^2 = 3.84$ where $\chi_{1,\alpha}^2$ is the $\alpha$th percentile of a chi-squared distribution with degree of freedom 1. Table \ref{Tab-Cov Invariance} reports the Wald statistics for the null hypotheses $H_{0,kj}$ $(k,j=1,2,3)$. For all three covariates, we find that there is no statistical evidence that the regression model varies over time at 5\% level. These empirical checks suggest that there is no significant evidence against \eqref{eq-model OREC}. 

\newpage

\begin{table}[!htp]
\renewcommand{\arraystretch}{1.2} \centering
\setlength{\tabcolsep}{4pt}
\begin{tabular}{|cc|ccc|}
\hline
\multicolumn{2}{|c|}{\multirow{2}{*}{\begin{tabular}[c]{@{}c@{}}Wald statistic for \\ $H_{0,kj}: \beta_{0kj}=\beta_{1kj}$\end{tabular}}} & \multicolumn{3}{c|}{$j$}                                                                                          \\ \cline{3-5} 
\multicolumn{2}{|c|}{}                                                                                                              & \multicolumn{1}{c|}{1 (Log population)} & \multicolumn{1}{c|}{2 (Population density)} & 3 (Proportion of females) \\ \hline
\multicolumn{1}{|c|}{\multirow{3}{*}{$k$}}                   & 0 (intercept)                   & \multicolumn{1}{c|}{0.680} & \multicolumn{1}{c|}{0.332} & 0.226 \\ \cline{2-5} 
\multicolumn{1}{|c|}{}                                       & 1 $(Y_t)$                       & \multicolumn{1}{c|}{1.281} & \multicolumn{1}{c|}{0.644} & 0.062 \\ \cline{2-5} 
\multicolumn{1}{|c|}{}                                       & 2 $(Y_t^2)$                     & \multicolumn{1}{c|}{1.719} & \multicolumn{1}{c|}{0.919} & 0.007 \\ \hline
\end{tabular}
\caption{\footnotesize Wald Statistics of the Null Hypotheses. Each row corresponds to the null hypotheses associated with the intercept term, linear term $(Y_t)$, and quadratic term $(Y_t^2)$, respectively. Each column corresponds to the null hypotheses associated with one of the three covariates (log population, population density, and proportion of females), respectively.}
\label{Tab-Cov Invariance} 
\end{table}

\newpage

\section{Proof of the Results} \label{sec:supp:Ident}

We first prove \eqref{eq-OR_representation}, which is restated below:
\begin{align*}
E\left( \potY{0}{1}  \cond A=1 \right)
=
E\left(  \potY{0}{1}   |A=1\right)  =
E\left[  \frac{E\left[
\obsY{1}  \exp\left\{  \beta_{0}\left(  \obsY{1} , X  \right)
\right\}  |A=0,X\right]  }{E\left[  \exp\left\{  \beta_{0}\left(  \obsY{1} , X
\right)  \right\}  |A=0,X\right]  } \Bigg| A=1 \right]
\end{align*}
The two terms of both hand sides are the same as follows:
\begin{align*}
& E \left( \potY{0}{1}  \cond A=1 \right)
\\
& = 
E \left\{ E \left( \potY{0}{1}  \cond A=1, X \right) \cond A=1 \right\}
\\
& =
E \left\{
\int y \cdot f \big( \potY{0}{1}=y \cond A=1 , X \big) \, dy \, \bigg| \, A=1 \right\}
\\
& = 
E \left[
\frac{ \int y \cdot h_1(y,X) \exp \big\{ \beta_0(y,X) \big\} \, dy }
{  \int h_1(y,X) \exp \big\{ \beta_0(y,X) \big\} \, dy }
\, \bigg| \, A=1 \right]
\\
& = 
E \left[
\Bigg[ 
\frac{ \int \big[ \exp \big\{ \beta_0(y,X) \big\} \big] \cdot h_1(y,X)  \, dy  }
{  \int h_1(y,X)  \, dy  }
\Bigg]^{-1}
\Bigg[
\frac{ \int \big[ y \cdot \exp \big\{ \beta_0(y,X) \big\} \big] \cdot h_1(y,X)  \, dy }
{  \int h_1(y,X)  \, dy  }
\Bigg] \, \Bigg| \, A=1 \right]
\\
& = 
E \left[
\frac{E\left[   \obsY{1} \exp\left\{  \beta_{0}\left(  \obsY{1},X  \right)  \right\} |A=0,X\right] }{E\left[  \exp\left\{  \beta_{0} \left( \obsY{1},X \right)  \right\}  |A=0,X\right] }
\bigg| A=1 \right]
\end{align*}
The first line holds from the law of iterated expectation. The second line holds from the definition of conditional expectation. The third line holds from \eqref{definition:h}. The fourth line holds from simple algebra. The last line holds from the definition of conditional expectation. Here, we remark that 
\begin{align}       \label{eq-xi}
\frac{E\left[   \obsY{1} \exp\left\{  \beta_{0}\left(  \obsY{1},X  \right)  \right\} |A=0,X\right] }{E\left[ \exp\left\{  \beta_{0} \left( \obsY{1},X \right)  \right\}  |A=0,X\right] }
=
E \left( \potY{0}{1}  \cond A=1, X \right) 
\end{align}

Next, we prove \eqref{eq-IPW_representation}, which suffices to show the following:
\begin{align*}
E\left(   \potY{0}{1}  \cond A=1\right)
= \frac{E\left\{  \left(  1-A\right)
\obsY{1}  \frac{\pi_{1}\left(  \obsY{1}, X  \right)  }%
{1-\pi_{1}\left( \obsY{1}, X \right)  }\right\}  }{E\left\{  \left(
1-A\right)  \frac{\pi_{1}\left( \obsY{1}, X \right)  }{1-\pi
_{1}\left( \obsY{1}, X \right)  }\right\}  }
=
\frac{E\left[ \left(  1-A\right)  Y_{1} \exp\left\{ (1,X\T)\eta_1 +  \alpha\T S_{0}( Y_1,X )  \right\} \right] }{E\left[ \left(  1-A\right)  \exp\left\{ (1,X\T)\eta_1 +  \alpha\T S_{0}( Y_1,X )  \right\} \right]  }
\end{align*}
The second identity is trivial from the parametrization of $\pi_1(Y_1,X)$. To show this, we establish the following intermediate result:
\begin{align*}
E\left\{  \left(  1-A\right)
\obsY{1}  \frac{\pi_{1}\left(  \obsY{1}, X  \right)  }%
{1-\pi_{1}\left( \obsY{1}, X \right)  }\right\}
& =
E\left\{  \left(  1-A\right)
\potY{1}{0}  \frac{\pi_{1}\left(  \potY{1}{0}, X  \right)  }%
{1-\pi_{1}\left( \potY{1}{0}, X \right)  }\right\}
\\
& =
E\left\{ \potY{0}{1}  \frac{\pi_{1}\left(  \potY{0}{1}, X  \right)  }%
{1-\pi_{1}\left( \potY{0}{1}, X \right)  } E \left(  1-A
\big| \potY{0}{1}, X \right) \right\}
\\
& =
E\left\{ \potY{0}{1} \pi_{1}\left(  \potY{0}{1}, X  \right) \right\}
\\
& =
E\left\{ \potY{0}{1} E \left( A
\big| \potY{0}{1}, X \right) \right\}
\\
& =
E\left( A \potY{0}{1} \right) 
\end{align*}
The first identity is from the consistency $\obsY{1}=\potY{0}{1}$ if $A=0$. The second identity is from the law of iterated expectation. The third and fourth identities are from the definition of $\pi_1$. The last identity is again from the law of iterated expectation. From analogous algebra, we also establish
\begin{align*}
E\left\{  \left(  1-A\right)  \frac{\pi_{1}\left(  \obsY{1}, X  \right)  }%
{1-\pi_{1}\left( \obsY{1}, X \right)  }\right\}
=
E\left( A \right) 
\end{align*}
Therefore, we establish  \eqref{eq-IPW_representation}:
\begin{align*}
E\left(   \potY{0}{1}  \cond A=1\right)
=
\frac{E(A \potY{0}{1} ) }{E(A)}
=
\frac{E\left\{  \left(  1-A\right)
\obsY{1}  \frac{\pi_{1}\left(  \obsY{1}, X  \right)  }%
{1-\pi_{1}\left( \obsY{1}, X \right)  }\right\}  }{E\left\{  \left(
1-A\right)  \frac{\pi_{1}\left( \obsY{1}, X \right)  }{1-\pi
_{1}\left( \obsY{1}, X \right)  }\right\}  }
\end{align*}

Lastly, we show that \eqref{eq-ATT0-DR} is equal to $E(\potY{0}{1} \cond A=1)$ either $\pi_{1}^{\dagger} (0,X) = \pi_{1} (0,X)$ or $\xi^{\dagger} = \xi$ where $\xi^{\dagger}$ and $\xi$ are shorthands for
\begin{align*}
\xi^{\dagger}(X)
= \frac{E^{\dagger}\left[  \obsY{1}  \exp\left\{  \alpha
_{0}\T S_{0}( \obsY{1} , X  )\right\}  |A=0,X\right]  }{E^{\dagger
}\left[  \exp\left\{  \alpha_{0}\T S_{0}(\obsY{1} , X  )\right\}
|A=0,X\right]}
\ , &&
\xi(X)=\frac{E\left[  \obsY{1}  \exp\left\{  \alpha
_{0}\T S_{0}( \obsY{1} , X  )\right\}  |A=0,X\right]  }{E\left[  \exp\left\{  \alpha_{0}\T S_{0}(\obsY{1} , X  )\right\}
|A=0,X\right]}
\end{align*}
Under the new notation, \eqref{eq-ATT0-DR} is simplified as
\begin{align*}
& E \left[  \frac{\left(  1-A\right)  }{\Pr\left(  A=1\right)  }\frac{\pi
_{1}^{\dagger}\left(  \obsY{1}  ,X\right)  }{  1-\pi_{1}^{\dagger
} ( \obsY{1} ,X )   } \Big\{  \obsY{1}
- \xi^{\dagger}(X) \Big\} + {\frac{A \xi^{\dagger}(X) }{\Pr(A=1)}} \right]
\end{align*}
First, we suppose $\pi_1^\dagger (0,X)=\pi_1 (0,X)$. Then, we find 
\begin{align*}
\pi_1^\dagger(y,x)
=
\exp \big\{ (1,x)\T  \eta_1^{\dagger} + \alpha_0\T  S_0(y,x) \big\}
=
\exp \big\{ (1,x)\T  \eta_1 + \alpha_0\T  S_0(y,x) \big\}
=
\pi_1(y,x)
\end{align*}
Consequently, we establish

\begin{align*}
& E  \left[  \frac{\left(  1-A\right)  }{\Pr\left(  A=1\right)  }\frac{\pi_{1}^{\dagger}\left(  \obsY{1}  ,X\right)  }{ 
1-\pi_{1}^{\dagger}\left( \obsY{1} ,X \right)  } \Big\{  \obsY{1}
- \xi^{\dagger}(X) \Big\} + {\frac{A \xi^{\dagger}(X) }{\Pr(A=1)}} \right]
\\
&
=
E\left[  \frac{\left(  1-A\right)  }{\Pr\left(  A=1\right)  }
\frac{\pi_{1} \left(  \potY{0}{1}  ,X\right)  }{  1-\pi_{1} \left( \potY{0}{1} ,X \right)  } 
\Big\{  \potY{0}{1} - \xi^{\dagger}(X) \Big\} 
+ {\frac{A \xi^{\dagger}(X) }{\Pr(A=1)}} \right]
\\
&
=
E\left[  \frac{ E\left(  1-A \cond \potY{0}{1},X \right)  }{\Pr\left(  A=1\right)  }
\frac{\pi_{1} \left(  \potY{0}{1}  ,X\right)  }{  1-\pi_{1} \left( \potY{0}{1} ,X \right) } 
\Big\{  \potY{0}{1} - \xi^{\dagger}(X) \Big\} 
+ {\frac{A \xi^{\dagger}(X) }{\Pr(A=1)}} \right]
\\
&
=
E\left[  \frac{ E\left(  A \cond \potY{0}{1},X \right)  }{\Pr\left(  A=1\right)  } 
\Big\{  \potY{0}{1} - \xi^{\dagger}(X) \Big\} 
+ {\frac{A \xi^{\dagger}(X) }{\Pr(A=1)}} \right]
\\
&
=
E\left\{  \frac{  A \potY{0}{1}  }{\Pr\left(  A=1\right)  } 
- {\frac{A \xi^{\dagger}(X) }{\Pr(A=1)}} 
+ {\frac{A \xi^{\dagger}(X) }{\Pr(A=1)}} \right\}
\\
& = E \left( \potY{0}{1} \cond A=1 \right)
\end{align*}
The second line is from $\pi_1^{\dagger}=\pi_1$ and the consistency $\obsY{1}=\potY{0}{1}$ if $A=0$. The third line is from the law of iterated expectation. The fourth and fifth lines are from the definition of $\pi_1$. The sixth line is again from the law of iterated expectation. The last line is trivial.

Second, we suppose $\xi^\dagger=\xi$. From \eqref{eq-xi}, $\xi(X) = E \big( \potY{0}{1} \cond A=1,X \big)$. Additionally, we find
\begin{align*}
& 
\frac{ E\left(  1-A \cond \potY{0}{1},X \right) }{ E\left(  A \cond \potY{0}{1},X \right) }
\frac{\pi_{1}^{\dagger}\left(  \potY{0}{1}  ,X\right)  }{ 1-\pi_{1}^{\dagger}\left( \potY{0}{1} ,X  \right)  } 
\\
& =
\frac{ 1-\pi_1(\potY{0}{1},X) }{ \pi_1(\potY{0}{1},X) }
\frac{\pi_{1}^{\dagger}\left(  \potY{0}{1}  ,X\right)  }{  1-\pi_{1}^{\dagger}\left( \potY{0}{1} ,X  \right)   } 
\\
& =
\exp
\left\{
-(1,x\T )\eta_1 - \alpha_0 S_0(\potY{0}{1},X)
+
(1,x\T )\eta_1^{\dagger} + \alpha_0 S_0(\potY{0}{1},X)
\right\}
\\
& =
\exp
\big\{ (1,x\T ) (\eta_1^{\dagger} - \eta_1 ) \big\}
\end{align*}
Returning back to \eqref{eq-ATT0-DR}, we obtain
\begin{align*}
& E\left[  \frac{\left(  1-A\right)  }{\Pr\left(  A=1\right)  }
\frac{\pi_{1}^{\dagger}\left(  \obsY{1}  ,X\right)  }{  1-\pi_{1}^{\dagger}\left( \obsY{1} ,X  \right)  } 
\Big\{  \obsY{1}- \xi^{\dagger}(X) \Big\} + {\frac{A \xi^{\dagger}(X) }{\Pr(A=1)}} \right]
\\
&
=
E\left[  \frac{\left(  1-A\right)  }{\Pr\left(  A=1\right)  }
\frac{\pi_{1}^{\dagger}\left(  \potY{0}{1}  ,X\right)  }{ 1-\pi_{1}^{\dagger}\left( \potY{0}{1} ,X \right)  } 
\Big\{  \potY{0}{1} - \xi(X) \Big\} 
+ {\frac{A \xi (X) }{\Pr(A=1)}} \right]
\\
&
=
E\left[  \frac{ E\left(  1-A \cond \potY{0}{1},X \right)  }{\Pr\left(  A=1\right)  }
\frac{\pi_{1}^{\dagger}\left(  \potY{0}{1}  ,X\right)  }{  1-\pi_{1}^{\dagger}\left( \potY{0}{1} ,X  \right)  } 
\Big\{  \potY{0}{1} - \xi (X) \Big\} 
+ {\frac{A \xi (X) }{\Pr(A=1)}} \right]
\\
&
=
E\Bigg[  \frac{ E\left(  A \cond \potY{0}{1},X \right)  }{\Pr\left(  A=1\right)  } 
\underbrace{ 
\frac{ E\left(  1-A \cond \potY{0}{1},X \right) }{ E\left(  A \cond \potY{0}{1},X \right) }
\frac{\pi_{1}^{\dagger}\left(  \potY{0}{1}  ,X\right)  }{  1-\pi_{1}^{\dagger}\left( \potY{0}{1} ,X  \right)  } 
}_{=: \exp ( (1,X\T ) (\eta_1^{\dagger} - \eta_1 ) ) }
\Big\{  \potY{0}{1} - \xi (X) \Big\} 
+ {\frac{A \xi (X) }{\Pr(A=1)}} \Bigg]
\\
&
=
E\left[  \frac{  \exp \{ (1,X\T ) (\eta_1^{\dagger} - \eta_1 ) \} A  \big\{  \potY{0}{1} - \xi(X) \big\}  }{\Pr\left(  A=1\right)  } 
+ {\frac{A \xi(X) }{\Pr(A=1)}} \right]
\\
&
=
E\left[   E \big[ \exp \{ (1,X\T ) (\eta_1^{\dagger} - \eta_1 ) \} \big\{ \potY{0}{1} - \xi(X) \big\} \cond A=1 \big]
+  E\big\{  \xi(X) \cond A=1 \big\} \right]
\\
&
=
E \Big[   E \big[ \exp \{ (1,X\T ) (\eta_1^{\dagger} - \eta_1 ) \} 
\big\{ \underbrace{ E\big( \potY{0}{1} \cond A=1,X \big) - \xi(X)}_{=0} \big\} \cond A=1 \big]
\Big]
\\
& \hspace{2cm} + E \left[ E\big\{  E(  \potY{0}{1} \cond A=1,X ) \cond A=1 \big\}  \right]
\\
& = E \left( \potY{0}{1} \cond A=1 \right)
\end{align*}
The second line is from $\xi^{\dagger}=\xi$ and the consistency $\obsY{1}=\potY{0}{1}$ if $A=0$. The third line is from the law of iterated expectation. The fourth line is straightforward from the established result above. The rest of the results are from the law of iterated expectation. This concludes the proof.

\newpage

\bibliographystyle{apa}
\bibliography{DiD.bib}

\end{document}